\documentclass[preprint]{elsarticle}
\usepackage[
  a4paper,
  left=20mm,
  right=20mm,
  top=25mm,
  bottom=25mm
]{geometry}
\usepackage{amsmath}
\usepackage{amsfonts}
\usepackage{amssymb}
\usepackage{amsthm}
\usepackage{array}
\usepackage{booktabs}
\usepackage{subfig}
\usepackage{tabularx}
\usepackage{comment}
\usepackage{xurl}
\usepackage{hyperref}

\usepackage{longtable}
\usepackage{array}
\usepackage{ragged2e}

\newcolumntype{P}[1]{%
  >{\RaggedRight\arraybackslash}p{#1}%
}

\journal{Computer Networks}

\theoremstyle{definition}
\newtheorem{example}{Example}[section]

\begin{document}

\begin{frontmatter}

\title{The Life of a Token: From Words to Bits on the Wire}

\author[cnam]{Davide Avesani\corref{cor1}}
\ead{davide.avesani@cnam.fr}

\author[cnam]{Pengwenlong Gu}
\ead{pengwenlong.gu@cnam.fr}

\author[cnam]{Sotiris Skaperas}
\ead{sotiris.skaperas@cnam.fr}

\author[cnam,neurotel]{Stefano Secci}
\ead{stefano.secci@cnam.fr}

\cortext[cor1]{Corresponding author}

\begin{abstract}
Large Language Models (LLMs) transform vast collections of unstructured
text into semantic patterns used for language generation and reasoning
tasks. Behind their ease of use lies a complex process: words become
tokens, tokens become vectors, and vectors ultimately give rise to streams
of bits that flow through High-Performance Computing (HPC) systems. As
modern LLMs grow to billions or trillions of parameters, this path
increasingly unfolds across thousands of interconnected accelerators,
making the underlying communication fabric a critical and often opaque
component of model training. This tutorial follows the journey from words to network traffic and
explains how language is translated into communication flows within HPC
training systems. Using concrete examples from Dante's \emph{Divine
Comedy}, we illustrate how model architecture, tokenization, embeddings,
and parallelization strategies shape the volume, structure, and timing of
data exchanged across the network. We combine architectural analysis,
analytical traffic models, and numerical examples to characterize the
communication requirements of LLM training. Our goal is to demystify how
words travel across the network and provide practical insights into the
network capabilities required to support the journey from text to a
trained model.
\end{abstract}

\begin{keyword}
large language models \sep
distributed training \sep
collective communication \sep
communication traffic modeling \sep
high-performance computing
\end{keyword}

\end{frontmatter}

\section{Introduction: The journey begins}
\label{sec:introduction}
In the opening lines of La Divina Commedia, Dante Alighieri finds himself in a dark forest~\cite{divinaCommedia}, poised at the threshold of transformation. Large Language Model (LLM) systems begin in a similarly opaque realm: a vast, unstructured forest of text.
Through training, this text is transformed into statistical and semantic representations that enable language generation, reasoning, and generalization across unseen tasks~\cite{radford2019language}. Beneath the natural language ability of LLMs lies a concrete systems process. Words are transformed into tokens, tokens into vector representations, vectors into activations and gradients, and these \lq tensors\rq \ ultimately into streams of bits exchanged across large scale distributed infrastructures.

This tutorial examines this transformation from a communication oriented perspective. Rather than focusing on model accuracy or architectural optimization, we focus on how the operations performed during LLM training translate into data movement among accelerators. Selected passages from La Divina Commedia are used as a running example to ground the discussion. By following specific fragments of the poem through tokenization, distribution, and processing, we connect abstract training concepts to concrete data flows within the infrastructure. This narrative anchors technical concepts to a tangible reference point, enabling an intuitive and structured explanation of the computational and communication mechanisms involved.

Tracing the life of a token has become increasingly challenging as it requires the understanding of both the model that processes it, and the infrastructure that makes such processing possible at scale. Since the introduction of the Transformer architecture~\cite{vaswani2017attention}, improvements in LLM performance have been largely driven by increases in model size, training data, and computational budget. This trend is reflected in the progression from models with hundreds of millions of parameters, such as GPT-2, to hundreds of billions of parameters, as in GPT-3, and more recently to models approaching the trillion parameter scale. Scaling law studies further show that model performance follows predictable trends as model size, dataset size, and compute increase~\cite{kaplan2020scaling}. This scaling pushes training far beyond the computational and memory capacity of a single machine. Modern LLMs are therefore trained on large clusters composed of hundreds or thousands of specialized accelerators, typically Graphics or Tensor Processing Units  (GPUs or TPUs). For instance, GPT-3 was reportedly trained using more than 10,000 (NVIDIA V100) GPUs~\cite{alarcon2020gpt3}, while GPT-4 is estimated to have used on the order of tens of thousands of (NVIDIA A100) GPUs~\cite{tithi2025scaling}.

The distribution of training across such a large system considerably increases the complexity of tracing this journey. Moreover, in such large distributed training scenario, computation is no longer the only limiting factor. Model parameters, gradients, optimizer states, and intermediate activations must be exchanged across devices, making communication a central component of the training process. Several studies have shown that communication can account for a significant fraction of LLM training time and, depending on the specific scenario, may even become the main performance bottleneck ~\cite{cheng2024thorough,wang2019characterizing,li2023accelerating, hanindhito2024bandwidth,jiang2024megascale,narayanan2021efficient}. As a result, improving large scale LLM training requires not only faster accelerators, but also communication infrastructures that are optimized to support the traffic generated during training. Accurately characterizing LLM training traffic is therefore essential for designing infrastructures capable of supporting future training workloads, including scenarios in which accelerator clusters are provided as shared or on demand G/TPU as a Service 
platforms. 
\begin{figure}[t]
    \centering
    \includegraphics[width=\linewidth]{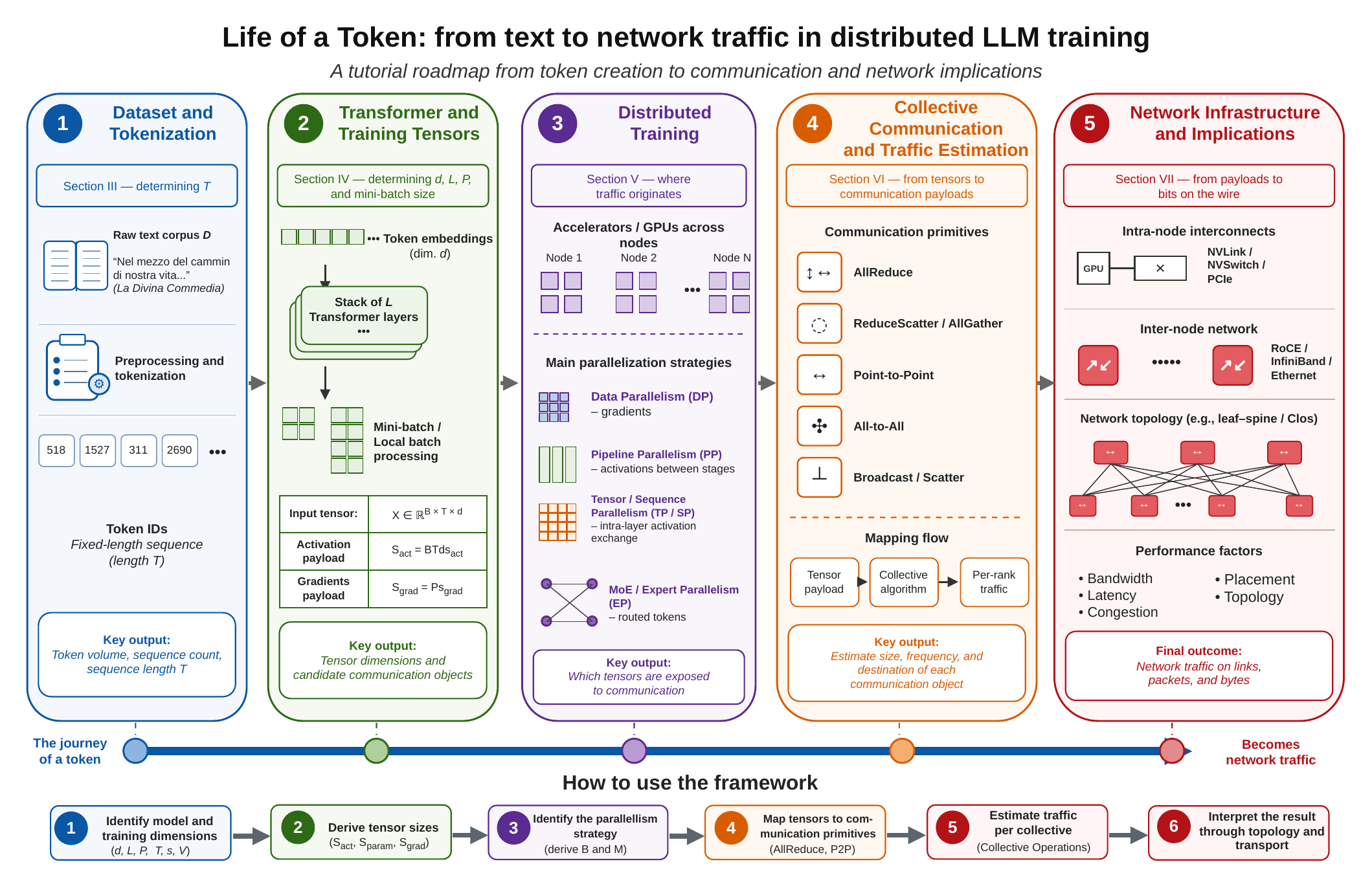}
    \caption{Illustration of the \textit{Life of a Token} framework. The figure traces the transformation of a token across the distributed LLM training stack, from raw text and tokenization, through model processing and parallel execution, to the network communication generated during large scale training.}
    \label{fig:the_life_of_a_token_framework}
\end{figure}
A complete account of this communication behavior is challenging because it is jointly shaped by multiple, tightly coupled factors: (i) dataset properties and partitioning, (ii) model architecture and scale, (iii) parallelization strategies and workload distribution, (iv) Collective Communication Operations (CCOs), and (v) network topology, interconnect technologies, and communication protocols. These factors are typically studied across separate communities, including machine learning, distributed systems, and networking. As a result, their combined effect on communication behavior remains difficult to capture, making accurate prediction and system-level optimization a nontrivial problem. To bridge these perspectives, we introduce the \lq Life of a Token\lq\ framework, illustrated in Fig.~\ref{fig:the_life_of_a_token_framework}, as the organizing abstraction of the tutorial: it  guides the reader through data processing during distributed LLM training, from raw text to tokens, Transformer tensors, and network traffic. Linking these stages provides a unified view of how decisions made at the data, model, and execution levels imply changes at the network-level.

\subsection{Tutorial objective and contributions}
The literature on LLMs has expanded rapidly in recent years, driven by significant industrial investment, continuous architectural innovation, and the growing scale of deployed models. At the same time, training these models across increasingly large and distributed accelerator clusters has made the network a critical component of the overall training system. Optimizing network performance is therefore essential, but doing so requires first characterizing the communication traffic generated during training. 

Although prior work provides detailed insights into specific aspects of LLMs and distributed training, the resulting body of knowledge remains fragmented across machine learning, distributed systems, and networking. As a result, translating high level model descriptions into concrete communication requirements for real world deployments remains challenging. This tutorial addresses this gap by providing a communication oriented framework for reasoning about how LLM training generates network traffic. We identify four limitations that motivate the need for such a tutorial.
\begin{enumerate}
\item \textbf{Limited connection between model design and infrastructure design.}
The literature provides extensive resources on the end-to-end development of LLMs, including widely recognized works such as~\cite{raschka2024build} and~\cite{alammar2024hands}. However, these contributions primarily focus on model architecture, training procedures, and implementation details, while providing limited insight into how model-level choices translate into infrastructure requirements. In particular, the impact of AI model parameters on communication volume and network requirements remains insufficiently exposed.

\item \textbf{Lack of systematic tools for communication modeling.}
Since the introduction of GPT-3~\cite{brown2020language}, LLM architectures have rapidly evolved, incorporating numerous optimizations and design variations. This diversity makes it difficult to consistently evaluate the communication implications of different configurations. A step by step methodology for estimating communication payloads and mapping them to network-level requirements is still largely missing.

\item \textbf{Fragmented analysis of communication mechanisms.}
Prior work has analyzed and demystified important components of the distributed training stack, including communication patterns in Transformer models~\cite{anthony2024demystifying}, collective communication protocols and algorithms~\cite{hu2025demystifying}, large scale parallel training systems~\cite{jiang2024megascale, narayanan2021efficient}, and production network fabrics for AI workloads~\cite{gangidi2024rdma}. However, these contributions are typically organized around individual layers of the stack, rather than around the end-to-end path from data to network traffic. A unified perspective that connects dataset processing, model architecture, tensor dimensions, parallelization choices, CCOs, and network infrastructure is still needed.

\item \textbf{Incomplete reporting of communication relevant training details.}
Technical reports on state of the art (SOTA) LLMs often emphasize architecture, dataset scale, context length, and benchmark performance~\cite{liu2024deepseek, guo2025deepseek, hui2024qwen2, touvron2023llama, anil2023palm, achiam2023gpt}. In contrast, details such as parallelization strategy, accelerator placement, interconnect technology, and network topology are often omitted or reported inconsistently. This makes it difficult to translate model level descriptions into communication payloads and network requirements. 

\end{enumerate}
Closest works take complementary but distinct perspectives. Duan et al.~\cite{duan2026efficient} survey distributed LLM training across infrastructure, parallelization, system optimization, and reliability, prioritizing comprehensive coverage rather than a continuous derivation of communication traffic. Liang et al.~\cite{liang2024communication} classify communication efficient techniques across algorithms, frameworks, and infrastructure, focusing primarily on mechanisms for reducing communication overhead rather than on how model and training choices generate that communication. Tazi et al.~\cite{tazi2025ultrascale} provide a practical tutorial with code and extensive scaling experiments, emphasizing how distributed training techniques are implemented and combined. Song et al.~\cite{song2026collective} instead follow collective operations through planning, execution, adaptation, and computation-communication coordination, after those operations have already been exposed by the workload. The \textit{Life of a Token} complements these works by beginning upstream, with raw text and a specified model and training configuration, and following a consistent derivation through tokenization, tensor construction, parallelization, collective invocations, and network traffic. Its contribution is not broader coverage or a new parallelization algorithm, but a reusable cross-layer derivation method. Fig.~\ref{fig:the_life_of_a_token_framework} summarizes the framework, formalized in Section~\ref{framework_overiew}.
The tutorial makes four contributions:
\begin{enumerate}

\item We present a token-centered, five-stage framework connecting dataset processing, Transformer tensors, distributed execution, collective communication, and network infrastructure.

\item We provide a step-by-step methodology for estimating the principal tensor payloads, communication operations, invocation frequencies, and per-accelerator and aggregate traffic associated with representative DP, PP, and TP configurations.

\item We connect logical payload estimates to representative collective algorithms and discuss how interconnect bandwidth, latency, topology, and accelerator placement influence communication behavior and training scalability.

\item We combine examples based on Dante's \textit{Divine Comedy} with a consistent GPT-2-like reference configuration to turn abstract model and training quantities into concrete payload and traffic estimates.

\end{enumerate}
Table~\ref{tab:positioning} summarizes these distinctions. The remainder follows the framework's five stages, while~\ref{app:extensions} provides additional derivations and extends the analysis to Mixture-of-Experts (MoE) models.
\begingroup
\scriptsize
\setlength{\tabcolsep}{3pt}
\renewcommand{\arraystretch}{0.95}

\begin{longtable}{@{}
  P{0.18\textwidth}
  P{0.25\textwidth}
  P{0.25\textwidth}
  P{0.24\textwidth}
@{}}

\caption{Positioning of this tutorial with respect to representative adjacent bodies of work.}
\label{tab:positioning}\\
\hline
\textbf{Adjacent line of work} &
\textbf{Representative references} &
\textbf{Main focus} &
\textbf{Difference from this tutorial} \\
\hline
\endfirsthead

\multicolumn{4}{c}{%
  \tablename\ \thetable{} -- continued from previous page
}\\
\hline
\textbf{Adjacent line of work} &
\textbf{Representative references} &
\textbf{Main focus} &
\textbf{Difference from this tutorial} \\
\hline
\endhead

\hline
\multicolumn{4}{r}{Continued on next page}\\
\endfoot

\hline
\endlastfoot

LLM foundations, architectures, and scaling &
Transformer and GPT-style models, scaling laws, and general LLM surveys
\cite{radford2019language,vaswani2017attention,kaplan2020scaling,
brown2020language,zhao2023survey,wang2025history} &
Explain the architectural principles of LLMs, including attention, Transformer blocks, decoder-only models, and the scaling of model size, data, and compute. &
Provide limited guidance on how architectural quantities translate into communication payloads and network requirements. \\
\hline

Datasets and tokenization &
Large-scale corpora, dataset construction, tokenization algorithms, and token-counting rules of thumb
\cite{liu2025datasets,commoncrawl,gao2020pile,gage1994new,
kudo2018sentencepiece,song2021fast,rajaraman2024toward,
OpenAITokensHelp,GoogleGeminiTokens} &
Describe how raw text is collected, cleaned, tokenized, and organized into token sequences for LLM training. &
Focus on data preparation but rarely follow how tokenization and sequence construction influence tensor dimensions and distributed communication payloads. \\
\hline

Distributed LLM training and parallelization &
Data, tensor, pipeline, sequence, and hybrid parallelism; memory optimization; distributed-training surveys, systems, and practical tutorials
\cite{duan2026efficient,tazi2025ultrascale,shoeybi2019megatron,
narayanan2021efficient,rasley2020deepspeed,rajbhandari2020zero,
zhao2023pytorch,huang2019gpipe,harlap2018pipedream,
li2021terapipe,qi2024zero,shazeer2018mesh,zheng2022alpa,
xu2021gspmd,ren2021zero,rajbhandari2021zero,wang2024zero++} &
Explain, implement, or optimize techniques for partitioning data, parameters, layers, activations, and computation across accelerators. &
Organize the problem around parallelization techniques, training systems, or implementation guidance rather than a continuous derivation from tokenized input to communication payloads and network requirements. \\
\hline

Collective communication and communication libraries &
Collective primitives, CCL implementations, collective synthesis, topology-aware algorithms, and collective-centric tutorials
\cite{song2026collective,nccl,nvidia_nccl_collectives,
weingram2023xccl,hu2025demystifying,cai2021synthesizing,
shah2023taccl,cowan2023mscclang,liu2024rethinking} &
Analyze operations such as AllReduce, ReduceScatter, AllGather, Broadcast, point-to-point exchange, and All-to-All, together with their planning and implementation over GPU clusters. &
Generally begin with already-defined messages or collective operations, leaving the upstream connection to tokenization, model tensors, and parallelization choices implicit. \\
\hline

AI cluster and data-center networking &
GPU interconnects, RDMA, InfiniBand, RoCE, NVLink/NVSwitch, TPU-scale systems, congestion control, and AI-oriented data-center fabrics
\cite{li2019evaluating,potluri2013efficient,pfister2001introduction,
kalia2016design,mittal2018revisiting,zhu2015congestion,
gangidi2024rdma,qian2024alibaba,jouppi2023tpu,
nvidia_spectrum_x_2024} &
Study the infrastructure that transports training traffic, including intra-node and inter-node interconnects, transport protocols, congestion mechanisms, and data-center topologies. &
Typically treat the communication workload as an input to network analysis rather than deriving it from model dimensions, training choices, and parallelization strategies. \\
\hline

Communication-oriented surveys, characterization, simulation, and traffic modeling &
Communication surveys, empirical characterization, bandwidth studies, training simulators, and topology-aware analyses
\cite{liang2024communication,anthony2024demystifying,
hanindhito2024bandwidth,tithi2025scaling,won2023astra,
wang2025simai,he2025efficient,liang2025lumos} &
Survey, measure, model, or simulate communication behavior in distributed Transformer and LLM training systems. &
Primarily produce taxonomies, platform-specific measurements, or simulator predictions rather than a single pedagogical procedure connecting raw text, tensor construction, collective invocations, and network traffic. \\
\hline

Recent large-model technical reports &
State-of-the-art LLM reports and large-scale model descriptions
\cite{achiam2023gpt,touvron2023llama,workshop2022bloom,
chowdhery2023palm,anil2023palm,team2024gemini,
bi2024deepseek,hui2024qwen2,guo2025deepseek} &
Report model architectures, training data, context lengths, scaling choices, and performance improvements for modern LLMs. &
Often provide limited visibility into the communication patterns, CCOs, network topology, and infrastructure assumptions required to train such models at scale. \\
\hline
\bottomrule
\end{longtable}
\endgroup

\section{The Life of a Token framework}
\label{framework_overiew}
This section formalizes the \textit{Life of a Token} framework that structures the rest of the tutorial. We first present it as a conceptual abstraction for following data across the training stack, and then describe how the same abstraction is used to reason about communication payloads. We then turn this abstraction into a step-by-step methodology for estimating communication payloads and conclude by defining the notation and modeling scope used in the rest of the paper. The notation, assumptions, and key quantities underlying the framework are introduced here and further developed in Section~\ref{transformer_chapter}, where we examine in detail how they arise from the LLM training pipeline.
\subsection{Framework overview}
At a high level, the framework provides the organizing abstraction used throughout the tutorial to connect the logical steps of LLM training with the communication traffic generated during distributed execution. In this view, the token acts as a conceptual thread: it starts as a discrete identifier produced by the tokenizer, becomes part of a dense activation tensor after the embedding layer, propagates through the Transformer stack, and eventually contributes to the tensors exchanged across accelerators during training. 
Fig.~\ref{fig:the_life_of_a_token_framework} summarizes the tutorial roadmap in five stages, which are formally introduced in Section~\ref{methodology}. We introduce the framework using a dense decoder-only Transformer as the reference path, since it cleanly exposes the main data transformations and communication mechanisms. More specialized architectures, especially MoE models, extend this baseline by adding communication stages through token routing. We use the dense case to build intuition and revisit MoE models later as a natural extension.

\subsection{Communication estimation methodology}
\label{methodology}
We now apply the five stage roadmap in Fig.~\ref{fig:the_life_of_a_token_framework} as an operational procedure for moving from the training input to the communication objects eventually observed by the network.

\begin{enumerate}
    \item Dataset and tokenization: the first stage determines how raw text is transformed into the token sequences processed during training. Starting from the training dataset, the text is cleaned, tokenized, and organized into fixed length sequences of length $T$. This stage determines the total number of tokens, the number of training sequences, and the sequence length used by the model.

    \item Transformer and training tensors: the second stage follows how token identifiers are mapped into dense embeddings and processed by the Transformer stack. For the dense decoder-only Transformer used as a reference, the main quantities are the vocabulary size $V$, the sequence length $T$, the hidden dimension $d$, the number of Transformer layers $L$, the number of trainable parameters $P$, the mini-batch size $B$, and the numerical precision $s$, expressed in bytes per value. 
    After the embedding layer, the input tensor has shape
    \begin{equation}
        X \in \mathbb{R}^{B \times T \times d}.
        \label{eq:activation_tensor_shape}
    \end{equation}
    The corresponding activation payload is
    \begin{equation}
        S_{\mathrm{act}} = B T d s_{\mathrm{act}}.
        \label{eq:activation_payload}
    \end{equation}
    Similarly, the parameter and gradient payloads can be approximated as
    \begin{subequations}
    \label{eq:model_payloads}
    
    \centering
    \begin{tabular}{@{}c@{\quad}c@{\qquad\qquad}c@{\quad}c@{}}
    
    \(\displaystyle S_{\theta}=P s_{\theta}\) &
    \refstepcounter{equation}
    \label{eq:parameter_payload}
    \textup{(\theequation)}
    &
    \(\displaystyle S_{\mathrm{grad}}=P s_{\mathrm{grad}}\) &
    \refstepcounter{equation}
    \label{eq:gradient_payload}
    \textup{(\theequation)}
    
    \end{tabular}
    
    \end{subequations}
    These quantities do not yet represent network traffic. They identify the candidate payloads that may become communication objects once training is distributed across multiple accelerators.

    \item Distributed training: the third stage considers how the training workload is partitioned across accelerators. Different parallelization strategies expose different tensors to communication, leading to distinct communication patterns and traffic characteristics. This stage therefore determines which tensors become communication objects.

    \item Collective communication and traffic estimation: the fourth stage maps the identified communication objects to abstract patterns and estimates their traffic. Depending on the parallelization strategy, tensors may be exchanged between replicas, passed between pipeline stages, or redistributed across devices within a layer. At this stage, the tensor determines the logical payload, while the communication pattern determines how that payload is exchanged across devices.

    \item Network infrastructure and implications: the fifth stage 
    maps the estimated payloads onto the underlying network infrastructure. Communication may remain within a node, traverse inter-node links, or cross multiple tiers of the data center network. Actual communication behavior depends on interconnect bandwidth, latency, routing, contention, congestion control, and topology aware placement. By identifying which tensors dominate communication, how often they are exchanged, and where they travel through the system, the framework eases reasoning about parallelization strategies, collective algorithms, accelerator placement, and network requirements.
\end{enumerate}
Overall, this methodology moves beyond a token-level view and frames the analysis as a cross layer reasoning tool. It links modeling and training decisions to the main communication objects, the ways they are exchanged, and the network resources they place under pressure.

\subsection{Notation and scope}
\label{notation_and_scope}
Table~\ref{tab:notation} summarizes the tutorial's notation.
We distinguish tensor payloads from network traffic: quantities such as $S_{\mathrm{act}}$, $S_{\theta}$, and $S_{\mathrm{grad}}$ identify the size of candidate communication objects, while the actual traffic observed on the network depends on the distributed training strategy, the collective communication algorithm, and the underlying topology. Low level implementation details, kernel level optimizations, memory allocation behavior, and framework specific scheduling choices are outside the main scope unless they directly affect communication patterns. 
Hence we want to emphasize the cross layer relationship between model configuration, tensor dimensions, communication primitives, and network-level behavior. 

\begin{table}[t]
\centering
\caption{Main notation used throughout the tutorial.}
\label{tab:notation}

\begin{tabularx}{\linewidth}{
  @{}
  l
  >{\raggedright\arraybackslash}X
  @{\hspace{1em}}
  l
  >{\raggedright\arraybackslash}X
  @{}
}
\toprule
\multicolumn{2}{c}{\textbf{Model and data}} &
\multicolumn{2}{c}{\textbf{Communication and storage}} \\
\cmidrule(lr){1-2}
\cmidrule(lr){3-4}
\textbf{Notation} & \textbf{Definition} &
\textbf{Notation} & \textbf{Definition} \\
\midrule
$B$
& Local mini-batch size.
& $C_{\mathrm{dev}}^{\mathcal{A}}(M,p_{\mathrm{cc}})$
& Per-accelerator communication volume. \\

$b_{\mu}$
& Pipeline micro-batch size.
& $C_{\mathrm{sys}}^{\mathcal{A}}(M,p_{\mathrm{cc}})$
& Aggregate communication volume. \\

$d$
& Model hidden size.
& $M$
& Logical communication payload in bytes. \\

$\mathcal{D}$
& Tokenized training dataset.
& $p_{\mathrm{cc}}$
& Number of communication participants. \\

$L$
& Number of Transformer layers.
& $s_{\mathrm{act}}$
& Bytes per activation value. \\

$m$
& Number of pipeline micro-batches.
& $s_{\mathrm{grad}}$
& Bytes per gradient value. \\

$P$
& Number of trainable parameters.
& $S_{\theta}=Ps_{\theta}$
& Full model parameter payload. \\

$T$
& Sequence length in tokens.
& \multicolumn{2}{l@{}}{} \\

$\mathcal{V}$
& Tokenizer vocabulary.
& \multicolumn{2}{l@{}}{} \\

$V=|\mathcal{V}|$
& Vocabulary size.
& \multicolumn{2}{l@{}}{} \\
\bottomrule
\end{tabularx}
\end{table}

\section{Dataset and tokenization: determining T}
\label{dataset_and_tokenization}
The training dataset plays a central role in LLM development, as its scale, quality, and diversity shape model behavior and performance. Modern LLM datasets aggregate heterogeneous sources, including web crawls (e.g., CommonCrawl~\cite{commoncrawl}, C4~\cite{raffel2020exploring}), books, public academic repositories (e.g., arXiv), code repositories (e.g., GitHub), Wikipedia, and large curated corpora such as The Pile~\cite{gao2020pile}. A comprehensive overview of such datasets is provided by Liu et al.~\cite{liu2025datasets}.

Once collected, the dataset undergoes \emph{preprocessing}, where raw data is cleaned, normalized, and structured for training~\cite{li2024datacomplm}. The resulting text is then tokenized, i.e., converted into a sequence of discrete units called \emph{tokens}, which provide the numerical input representation used by the LLM. 
Various techniques have been developed to perform tokenization, with subword based approaches such as Byte Pair Encoding (BPE)~\cite{gage1994new}, WordPiece~\cite{song2021fast}, and SentencePiece/Unigram LM~\cite{kudo2018sentencepiece} being the most widely adopted in modern LLMs. These methods balance vocabulary size and representation efficiency by decomposing text into frequent subword units. Recent work has also begun to formalize the theoretical role of tokenization in shaping model efficiency and behavior~\cite{rajaraman2024toward}. The tokenizer is typically trained separately prior to model training and remains fixed during optimization. Its design directly affects the number of tokens generated from raw text and, consequently, the amount of data that must be processed. 

\begin{figure}[t]
    \centering
    \includegraphics[
    width=\linewidth
    ]{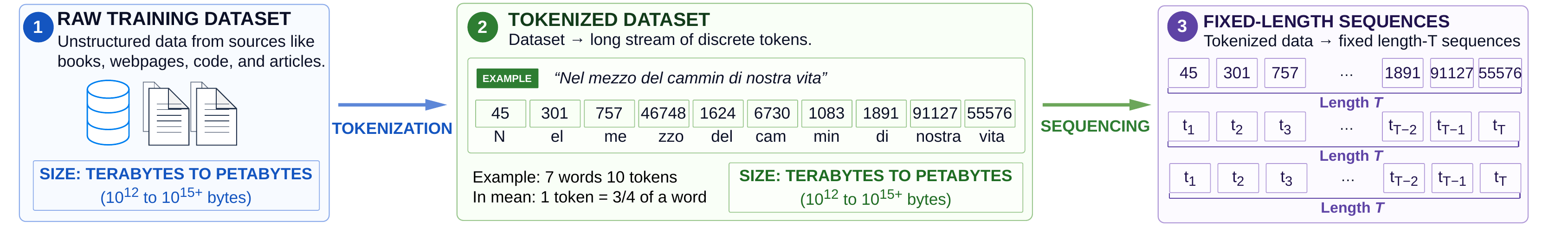}
    \caption{From dataset size to tokenization to sequence organization}
    \label{fig:tokenization}
\end{figure}

The tokenized dataset is then divided into fixed length \emph{sequences} of $T$ ordered token identifiers, with each sequence serving as one training sample. The sequence length $T$ is kept fixed during training to ensure a consistent input representation across samples, enabling efficient tensor based computation and stable optimization. The value of $T$ is kept constant within a training phase to obtain tensors with consistent dimensions. In practice, however, its value may be
increased across training stages to extend the supported context length~\cite{xiong2024effective,grattafiori2024llama}. The sequence length \(T\) is therefore an important model and training hyperparameter. Fig.~\ref{fig:tokenization} summarizes the transition from raw text to fixed length token sequences.

\subsection{Notation and dimensionality}
Dataset size typically scales with model capacity, ranging from tens of gigabytes (GB) for smaller models to multi-terabyte collections for large-scale systems, as illustrated by datasets such as Dolma and FineWeb~\cite{soldaini2024dolma,penedo2024fineweb}. During tokenization, the raw text is transformed into a sequence of discrete tokens drawn from a fixed vocabulary \(\mathcal{V}\), whose size is \(V=|\mathcal{V}|\). Let \(\mathcal{D}\) denote the tokenized training dataset. It can be represented as
$
\mathcal{D}=(x_1,x_2,\ldots,x_{|\mathcal{D}|}), x_i\in\mathcal{V},
$ where \(|\mathcal{D}|\) is the total number of tokens in the dataset. The vocabulary size \(V\) is a key design parameter in LLM engineering. It is typically fixed during model construction and influences both the expressiveness of the token representation and the number of tokens required to encode a given text. In practice, vocabulary sizes commonly range from tens of thousands to a few hundred thousand tokens, depending on the tokenizer design and the languages covered by the model~\cite{ali-etal-2024-tokenizer,liang2023xlm}.

The exact mapping from raw text to tokens depends on the tokenization algorithm and its implementation. As a commonly used rule of thumb for English text, one token corresponds on average to approximately four characters, or three quarters of a word~\cite{OpenAITokensHelp,GoogleGeminiTokens}. Because most characters in predominantly English text occupy one byte when encoded in UTF-8, this corresponds to an order-of-magnitude estimate of approximately four bytes of raw text per token. If \(S_{\mathrm{text}}\) denotes the size of the cleaned text in bytes and \(\beta_{\mathrm{tok}}\) the average number of raw-text bytes represented by one token, then
$
|\mathcal{D}| \approx
\frac{S_{\mathrm{text}}}{\beta_{\mathrm{tok}}},
$
where \(\beta_{\mathrm{tok}}\approx 4\) for English-dominated corpora.
The tokenized dataset is subsequently partitioned into \(N\) fixed-length training sequences,
$
S_i=[t_{i,1},t_{i,2},\ldots,t_{i,T}], i\in\{1,\ldots,N\},
$
where \(T\) is the sequence length and \(N\approx|\mathcal{D}|/T\). Depending on the preprocessing procedure, an incomplete final sequence may be padded, packed with tokens from another document, or discarded. In practice, typical values of \(T\) range from \(1024\) to \(4096\) tokens for many LLMs, while long-context models may use substantially larger values. Example~\ref{pb-token-scale} illustrates the scale of tokenized datasets and the resulting sequence counts encountered during LLM training.

\begin{example}[PB Corpus Token and Sequence Scale]
\label{pb-token-scale}
Consider a cleaned textual dataset of size \(S_{\mathrm{text}}=1\) PB, where
\(1~\mathrm{PB}=10^{15}\) bytes. Using the approximation
\(\beta_{tok}\approx 4\) bytes per token, the total number of tokens is
$
|\mathcal{D}|
\approx
\frac{S_{\mathrm{text}}}{\beta_{\mathrm{tok}}}
=
\frac{10^{15}}{4}
=
2.5\times 10^{14}.
$
Thus, a PB scale textual corpus corresponds to approximately $250$ trillion tokens. If the corpus is divided into fixed length sequences, the number of resulting sequences is approximately
$
N_{\mathrm{seq}}
\approx
\frac{|\mathcal{D}|}{T}.
$
For $T=1024$, this gives
$
N_{\mathrm{seq}}
\approx
\frac{2.5\times 10^{14}}{1024}
\approx
2.44\times 10^{11},
$
corresponding to about $244$ billion sequences.
For \(T=4096\), the number of sequences becomes
$
N_{\mathrm{seq}}
\approx
\frac{2.5\times 10^{14}}{4096}
\approx
6.10\times 10^{10},
$
corresponding to about $61$ billion sequences.
\end{example}
To provide an intuition of such scale, consider \textit{La Divina Commedia}, which contains approximately \(100{,}000\) words~\cite{DivinaCommediaTypesTokens}. Using the approximation that one word corresponds to about \(1.25\) tokens, the poem contains approximately
$
|\mathcal{D}|
\approx
100{,}000 \times 1.25
=
125{,}000 \ \mathrm{tokens}.
$
If the text is divided into sequences of length \(T=1024\), the number of sequences is approximately 123. Compared with the petabyte scale corpus discussed above, which contains approximately \(2.5\times 10^{14}\) tokens, the tokenized \textit{Divina Commedia} represents only $5\times 10^{-8}\%$ of the total dataset token volume. 

\section{From tokens to Transformer tensors}
\label{transformer_chapter}
The previous section described how raw text is converted into fixed-length token sequences of length $T$. In the Life of a Token framework, we next follow these sequences into the LLM where token-ID tensors are mapped to dense embeddings. After briefly introducing the Transformer architecture and the decoder-only reference model, this section traces how tokens are embedded and processed through the Transformer layers, establishing the main quantities used in the subsequent communication analysis, such as $V$, $T$, $d$, $L$, $P$, $B$,
$s_{\mathrm{act}}$ and $s_{\mathrm{grad}}$.

\subsection{From the original Transformer to decoder-only LLMs}
The Transformer architecture forms the core Neural Network (NN) design on which most modern LLMs are built. It was originally introduced as an encoder-decoder model for sequence-to-sequence modeling~\cite{vaswani2017attention}. It is composed of two main components: an encoder, which processes the input sequence, and a decoder, which generates the output sequence. Over time, this design has evolved into three main variants, each tailored to different types of tasks:

\begin{itemize}
\item \textbf{Encoder-decoder}: the 
original formulation, combining an encoder that processes the input and a decoder that generates the output. This structure is commonly used for sequence to sequence tasks such as machine translation and summarization (e.g., T5~\cite{raffel2020exploring}, BART~\cite{lewis2020bart}).

\item \textbf{Encoder-only}: a simplified version that retains only the encoder component. These models are primarily designed to understand and represent text, and are widely used for tasks such as classification, question answering, and information extraction (e.g., BERT~\cite{devlin2019bert}, RoBERTa~\cite{liu2019roberta}).

\item \textbf{Decoder-only}: a variant that uses only the decoder component to generate text, one token at a time. This architecture gained significant popularity with the introduction of the GPT-2 model~\cite{radford2019language} and gained widespread adoption with GPT-3~\cite{brown2020language}. It has now become the dominant choice for modern generative LLMs, including the GPT, Gemini, LLaMA, and Mistral model families~\cite{brown2020language,team2024gemini,grattafiori2024llama,jiang2023mistral}.
\end{itemize}

\subsection{Reference decoder-only Transformer architecture}
The decoder-only variant of the Transformer architecture~\cite{vaswani2017attention}, popularized by GPT-style models and in particular by GPT-2~\cite{radford2019language}, has become the foundation on which modern LLMs are built. As a reference, we adopt a simplified decoder-only Transformer inspired by the GPT-2 architecture~\cite{radford2019language}. Illustrations of the original encoder-decoder architecture and the decoder-only reference architecture are provided in the~\ref{app:architectural_context}. After GPT-2, many architectural variations have been proposed, but the decoder-only Transformer remains the dominant backbone of modern LLMs. A concise overview of these variants is provided in~\cite{raschka2024llm_gallery}. In this tutorial, however, we retain the proposed simplified structure throughout the paper for the following reasons:
\begin{itemize}
    \item It captures the core building blocks that remain common across most contemporary LLMs.
    \item It provides a clear and interpretable representation of how data propagates through the model, which is essential for analyzing communication patterns.
    \item Although modern architectures introduce optimizations, the fundamental communication behavior remains largely governed by the same objects, making this abstraction sufficient for our analysis.
\end{itemize}
To maintain clarity and avoid unnecessary complexity, we do not consider advanced architectural extensions here such as MoE, as they introduce specialized communication patterns. MoE architectures and their communication implications are examined in~\ref{app:moe}. The reference architecture is composed of the following components:
\begin{itemize}
    \item \textbf{Input embedding and positional encoding}: this layer transforms the processed tokens into vector representations. Each token is mapped to a dense vector of dimension $d$ through an embedding matrix of size $V\times d$, where $V=|\mathcal{V}|$ is the vocabulary size. The embedding matrix is part of the model parameters and is therefore learned jointly with the other model parameters. Positional information can also be added to the embedding representation through mechanisms such as rotary position embeddings~\cite{su2024roformer}.

    \item \textbf{$L$ stacked Transformer blocks}: each block takes a
    sequence of hidden representations as input and produces updated
    representations with the same overall tensor shape. The number of Transformer blocks $L$ is a design parameter defined during model construction and varies with the scale of the LLM. In practice, \(L\) can range from around a dozen to more than one hundred Transformer layers. For example, the OPT family ranges from \(12\) layers in OPT-125M to \(96\) layers in OPT-175B, while MT-NLG 530B uses \(105\) layers~\cite{zhang2022opt,smith2022using}.
    
    \item \textbf{Masked multi-head self-attention}: each of the L Transformer layers contains a \emph{Masked multi-head self-attention} block. This component represents the key innovation of the Transformer architecture; it allows each token to selectively relate to previous tokens in the sequence, building a context aware representation based on the surrounding elements.
    At a high level, this process is implemented through the scaled dot product attention mechanism, defined as:
    $
    \mathrm{Attention}(Q,K,\mathbf{V}{\mathrm{att}}) = \mathrm{softmax} \left( \frac{\mathbf{Q}\mathbf{K}^{\top}}{\sqrt{d_h}} \right) \mathbf{V}{\mathrm{att}},
    $
    where $\mathbf{Q}$, $\mathbf{K}$, and $\mathbf{V}{\mathrm{att}}$ are the query, key, and value tensors, respectively (the notation $\mathbf{V}_{\mathrm{att}}$ is used to avoid confusion with the vocabulary size $V$).
    To increase modeling flexibility, this operation is performed in parallel across multiple attention heads, allowing the model to capture different types of relationships within the sequence. 
    
    \item \textbf{Feed Forward Network (FFN)}: this is the second component inside each Transformer layer, which applies a transformation to each position separately, refining the representation obtained from the attention layer. At a high level, the FFN consists of two linear transformations separated by a nonlinear activation function, and can be expressed as:
    $ 
    \operatorname{FFN}(\mathbf{h}) = 
    \mathbf{W}_2\, \sigma\!\left( \mathbf{W}_1\mathbf{h}+\mathbf{b}_1 \right) + \mathbf{b}_2, 
    $
    where $\mathbf{h}\in\mathbb{R}^{d}$ is the representation of one token, $\mathbf{W}_1$ projects it from the hidden dimension $d$ to an intermediate dimension $d_{\mathrm{ff}}$, and $\mathbf{W}_2$ projects it back from $d_{\mathrm{ff}}$ to $d$. The vectors $\mathbf{b}_1$ and $\mathbf{b}_2$ are the corresponding bias terms. The function $\sigma(\cdot)$ denotes a nonlinear activation function, such as the Gaussian Error Linear Unit (GeLU)\footnote{GeLU is a smooth nonlinear activation function defined as $\operatorname{GELU}(x)=x\Phi(x)$, where $\Phi(x)$ is the cumulative distribution function of the standard normal distribution. Other Transformer architectures may use different activation functions or gated variants, such as ReLU, SiLU, or SwiGLU.} used in the proposed GPT-2 architecture, although the specific activation function depends on the model architecture.
    This operation temporarily expands the dimensionality of the representation (typically to a multiple of $d$) before projecting it back to the original size, allowing the model to capture more complex patterns.
    It is important to note that the FFN preserves the overall dimensionality of the data while further transforming each token representation before passing it to the next layer.

    \item \textbf{Output layer (linear projection and softmax)}: the final stage of the model transforms the output representations into predictions over the vocabulary. Each token representation, of dimension $d$, is projected onto the vocabulary space through a linear transformation defined by a matrix of size $V \times d$, where $V=|\mathcal{V}|$. This operation produces a vector of values known as \emph{logits}, which represent unnormalized scores for each possible token in the vocabulary. A softmax function\footnote{For a logit vector $\mathbf{z}\in\mathbb{R}^{V}$, softmax is defined as $\operatorname{softmax}(\mathbf{z})_i =\exp(z_i)\big/\sum_{j=1}^{V}\exp(z_j)$, producing nonnegative values that sum to one.} is then applied to convert these scores into a probability distribution over the vocabulary which are used to compute the cross entropy loss against the target token at each position. Due to the large vocabulary size, this projection is computationally and memory intensive, as it requires producing a vector of $V$ logits for each token position.
\end{itemize}
Several recent works further optimize the attention mechanism itself. FlashAttention~\cite{dao2022flashattention} improves memory efficiency through optimized attention kernels while FlashAttention 2~\cite{dao2024flashattention} further improves parallelism and work partitioning on the GPU. Linear attention~\cite{katharopoulos2020transformers} approximates attention to improve scaling with sequence length. Other variants modify how keys and values are shared across heads, including multi-query attention~\cite{shazeer2019fast} and grouped-query attention~\cite{ainslie2023gqa}. A broader overview of efficient attention mechanisms is provided in~\cite{sun2025efficient}. Although these techniques modify the internal execution of the attention block, they leave the input and output tensor dimensions unchanged. Since only internal attention execution changes, we do not distinguish among these variants. The FFN block has also been the target of efficiency improvements. Sparse and MoE variants increase model capacity without increasing the active computation proportionally~\cite{shazeer2017outrageously,lepikhin2020gshard,fedus2022switch,du2022glam}. As mentioned above, we discuss MoE based architectures separately in the~\ref{app:moe}.
\subsection{Inside the input pipeline}
\begin{figure}[t]
    \centering
    \includegraphics[
    width=\linewidth,
    trim={0 30 0 30},
    clip
    ]{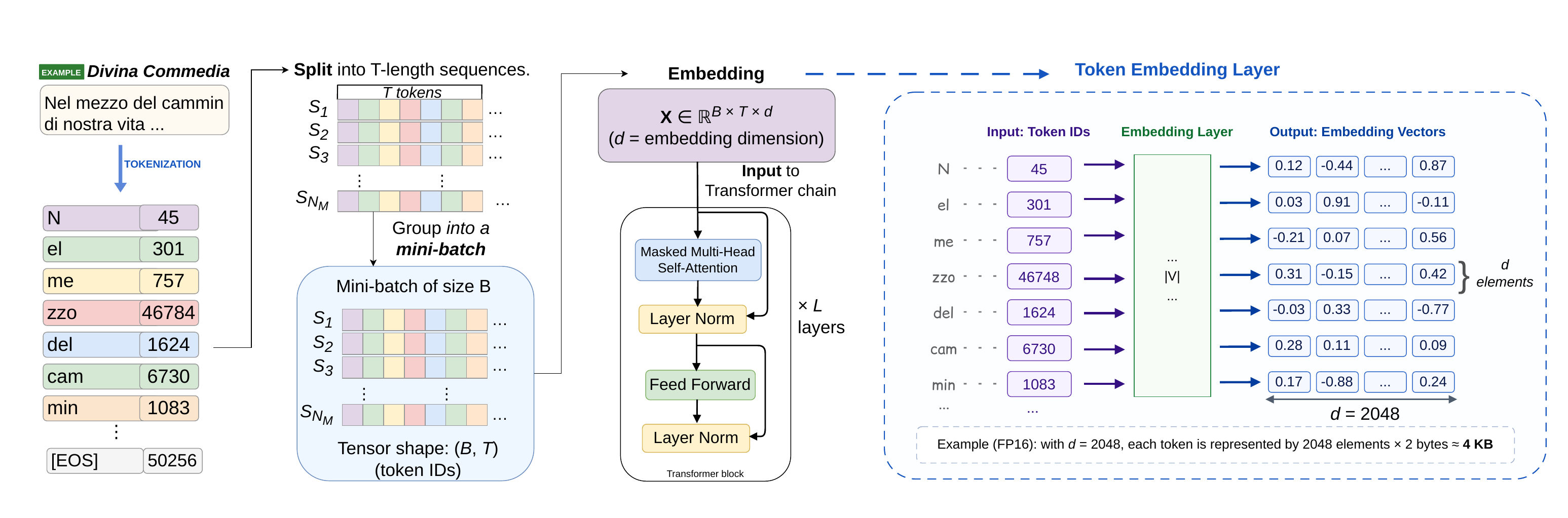}
    \caption{Illustration of the data flow during LLM training. Raw text from \textit{La Divina Commedia} is tokenized, divided into fixed length sequences, grouped into a mini batch of size $B$, and transformed into the input tensor $\mathbf{X} \in \mathbb{R}^{B \times T}$.}
    \label{fig:input_mini_batch}
\end{figure}

As introduced in Section~\ref{dataset_and_tokenization}, the basic
training sample processed by the model is a sequence of \(T\) token
identifiers. In practice, processing a single sequence at a time is generally inefficient, as it may underutilize the parallel computational resources available on modern GPUs. Instead, multiple sequences are grouped together into a tensor ($\mathbf{X}$), commonly referred to as a \emph{mini-batch}, containing $B$ sequences. The mini-batch size is chosen to maximize the throughput of the training system, i.e., the number of tokens processed per unit time. Intuitively, this corresponds to selecting the largest mini-batch that can fit within the available accelerator memory while avoiding out of memory conditions. Larger mini-batches generally improve hardware utilization, although the optimal value ultimately depends on both the model architecture and the available hardware resources. To summarize, the input tensor \(\mathbf{X}\) provided to the Transformer during training can be represented as
$
\mathbf{X} \in \{1,\dots,V\}^{B \times T},
$
where \(B\) denotes the mini-batch size, and \(T\) the sequence length. Fig.~\ref{fig:input_mini_batch} summarizes how raw text is transformed into tokenized sequences and subsequently aggregated into a mini-batch, which serves as the input to the LLM during training.

\subsection{The journey of the input tensor}
We now follow the journey of the input tensor \(\mathbf{X}\) through the Transformer model, focusing on the transformations it undergoes within the network. Understanding these transformations is essential for characterizing the communication and network traffic patterns discussed later in this tutorial. The first operation applied to the input tensor \(\mathbf{X}\) is the \emph{embedding} layer. This layer maps each token identifier to a dense vector representation of fixed dimension (d). In SOTA LLMs, the embedding dimension typically ranges from a few hundred to several thousand elements (e.g., \(d \in [768, 16{,}384]\)), with each element stored using numerical formats such as FP16, BF16, or FP32. As a result, each token is transformed from a single integer identifier into a high dimensional vector representation. Consequently, after the embedding layer, the input tensor assumes the form defined in~\eqref{eq:activation_tensor_shape} where each token is represented by a \(d\) dimensional embedding vector. Fig.~\ref{fig:input_mini_batch} illustrates this process, showing how token identifiers are mapped to their corresponding embedding vectors. A positional encoding is then typically added to the embedding representation to encode token ordering within the sequence, while preserving the same tensor dimensions. After this transformation, the tensor \(\mathbf{X}\) is commonly referred to as the \emph{input activation} or \emph{hidden state}.

The input activations are then processed by the first Transformer layer. In the decoder-only architecture considered here, each layer combines masked multi-head attention with a FFN. The tensor \(\mathbf{X}\) first enters the attention block, where token representations are updated using information from the rest of the sequence. The resulting tensor is then processed by the FFN, which further updates the embedding values. Throughout these operations, \textbf{the structure of the output tensor remains unchanged}: the number of sequences, the sequence length, and the embedding dimension are preserved. Consequently, the output of the Transformer layer is still represented as the tensor defined in Eq.~\eqref{eq:activation_tensor_shape} although the values contained in the embedding vectors have been updated.

\begin{figure}[t]
\centering
 \includegraphics[
    width=\linewidth
    ]{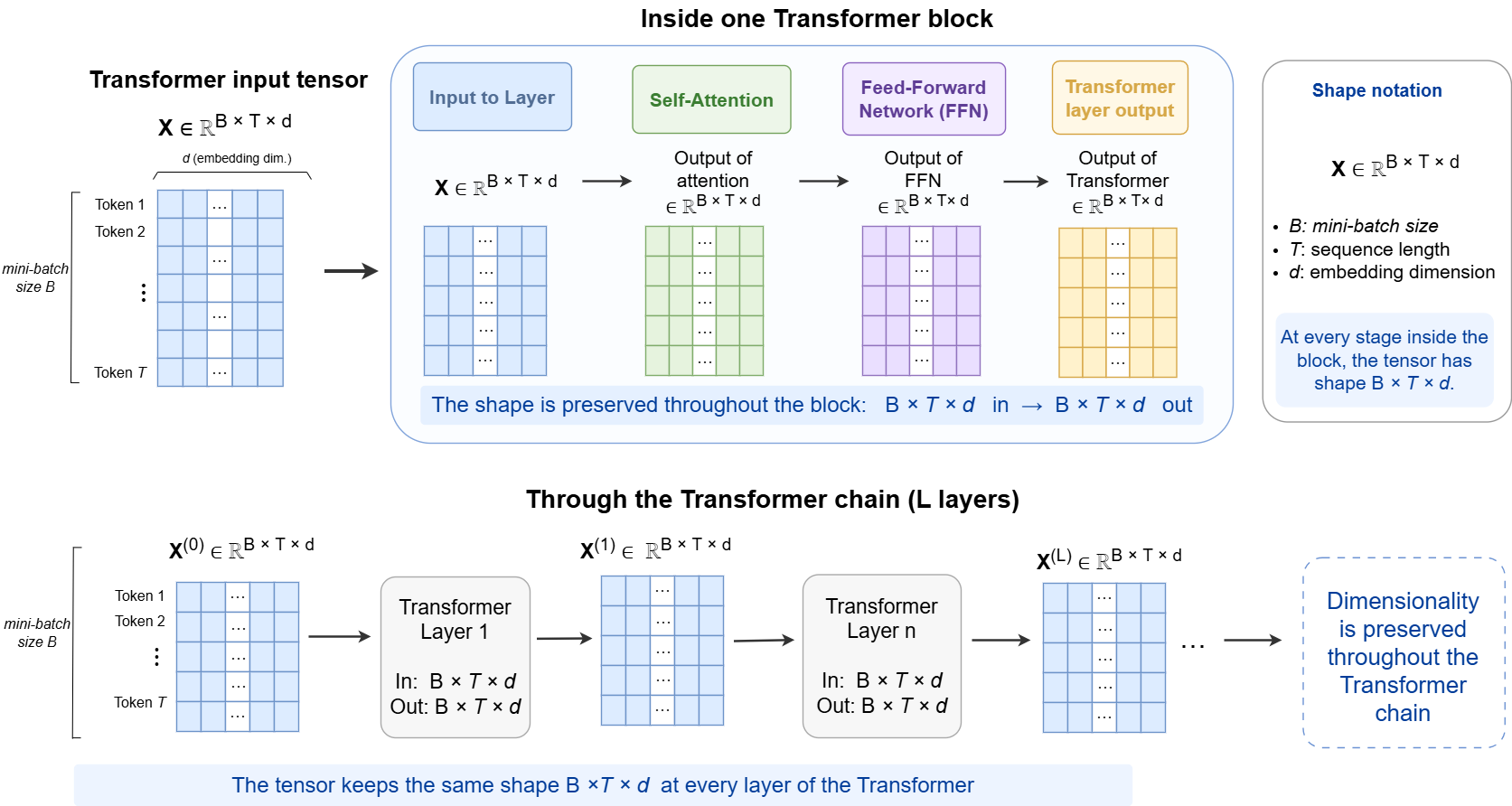}
\caption{Tensor shape preservation: $\mathbf{X} \in \mathbb{R}^{B \times T \times d}$ retains its shape across attention, FFN, and the Transformer chain.}
\label{fig:Forward_pass}
\end{figure}

The output tensor is then passed to the next Transformer layer, where the same process is repeated. Layer after layer, the embedding vectors are progressively refined as increasingly rich contextual information is incorporated into their representations. This process continues until the final Transformer layer is reached. Throughout the entire network, the tensor \textbf{retains the same shape}, with only the values of its embeddings being updated. Fig.~\ref{fig:Forward_pass} summarizes this process, illustrating how the embedding values are progressively updated while the tensor shape remains unchanged throughout the Transformer chain.
The first significant dimensionality change occurs only at the final projection layer. Specifically, the tensor \(\mathbf{X}\) is projected onto the vocabulary space, producing a tensor of logits
$
\mathbf{Z} \in \mathbb{R}^{B \times T \times V}.
$
Each vector of length ($V$) contains a score for every possible next token in the vocabulary. These scores are subsequently converted into probabilities through a softmax operation, yielding an output tensor of the same dimensions.

\subsection{The LLM training step}
We have followed the journey of an input mini-batch through the Transformer model and observed how its dimensionality remains unchanged throughout the network. However, this journey represents only one component of the overall training process. To correctly introduce the distributed training techniques discussed in the following sections, we first review the fundamental concepts and terminology associated with LLM training. A training step can be conceptually divided into four main phases:
\begin{enumerate}
\item  \textbf{Forward pass}: this phase corresponds to the process described in the previous section. The input tensor is first transformed into input activations by the embedding layer, and is subsequently processed by the sequence of Transformer layers. The final tensor representation is then projected onto the vocabulary space and used to predict the next token at each position of the sequence.
\item \textbf{BW pass}: Once the training loss has been computed, the model must determine how each parameter contributed to the prediction error. This information is captured by quantities known as \emph{gradients}, which indicate how the loss would change if a given parameter were modified. Backpropagation computes these gradients and propagates them from the output layer to the input layer. From a data-flow perspective, the backward (BW) pass mirrors the forward (FW) pass in reverse, propagating gradients rather than activations. These gradient tensors retain the overall dimensions of the corresponding activations as they traverse the Transformer layers.
\item \textbf{Gradient accumulation}: a mini-batch represents the amount of data that can be processed by the model at a given time and is typically constrained by the available accelerator memory. However, model parameters are not necessarily updated after every mini-batch. Instead, multiple mini-batches are often processed before completing a training step. The set of tokens processed before a parameter update is commonly referred to as the \emph {batch} (or \emph{global batch}): LLM training runs typically use global batch sizes of roughly $10^6$ to $10^7$ tokens per training step \cite{zhang2024does}. Since processing an entire global batch simultaneously is often infeasible, it is divided into multiple mini-batches, which are processed sequentially. The gradients computed during the BW pass of each mini-batch are accumulated, typically through summation, and are only used to update the model parameters once all mini-batches belonging to the global batch have been processed.
\item \textbf{Weight update}: After computing the gradients, the model parameters are updated using an optimization algorithm. The new parameters are obtained by adjusting the current weights based on the computed gradients, moving them in a direction that reduces the loss and improves the model's predictions. This update completes a training iteration, commonly referred to as a \emph{training step}.
\end{enumerate}
\subsection{Batch Scale and Single GPU Training}
To illustrate the scale of batches used in LLM training, consider \textit{La Divina Commedia}, whose approximately \(100{,}000\) words ~\cite{DivinaCommediaTypesTokens} correspond to roughly (125{,}000) tokens. For comparison, a model such as GPT-2 is commonly trained using batch sizes on the order of ($5 \times 10^5$) tokens. In this case, the entire \textit{Divina Commedia} would occupy only about one quarter of a single training batch. Considering instead GPT-3 175B, which uses a batch size of approximately ($3.2 \times 10^6$) tokens, the complete \textit{Divina Commedia} would represent only
$
\frac{125{,}000}{3.2 \times 10^6}\times 100
\approx
3.9\%
$
of a single batch. Equivalently, approximately 26 copies of the entire poem would be required to fill one GPT-3 training batch. Let us now follow a concrete mini-batch through the training pipeline and examine the dimensions of the input tensor \(\mathbf{X}\) and the outputs generated at each stage. We base this example on a modified version of the implementation of the GPT-2 model provided in~\cite{karpathy_build_nanogpt_train_gpt2}, and we report the tensor dimensions and throughput observed in the single GPU setup.

\begin{example}[Processing a mini-batch on a single GPU]
\label{single_gpu_mini_batch}

Consider running the GPT-2 model implementation with sequence length \(T=1024\) and hidden dimension \(d=768\), where activations are stored using FP16 precision (2 bytes per value), running on an NVIDIA L40S accelerator equipped with 48~GB of VRAM. A single input sequence of length 1024 is first transformed by the embedding layer into a tensor containing
$
1024 \times 768 = 786{,}432\ \text{FP16 values}
$.
The resulting activation tensor therefore occupies
$
1024 \times 768 \times 2
=
1{,}572{,}864\ \text{bytes}
\approx
1.5\ \text{MB}
$.
With this setup, processing a single sequence achieves a throughput of approximately \(65{,}000\) tokens/s. Loading the model and processing the sequence requires about 2~GB of GPU memory. The available GPU memory therefore allows multiple sequences to be processed simultaneously. In this setup, the maximum mini-batch size that could be accommodated was $B=32$ sequences. Consequently, the input tensor before the embedding layer contains $32 \times 1024 = 32{,}768\ \text{tokens}$ .
After the embedding layer, the tensor becomes
$
\mathbf{X}
\in
\mathbb{R}^{32 \times 1024 \times 768}
$
with a corresponding activation memory of
$
32 \times 1024 \times 768 \times 2
=
50{,}331{,}648\ \text{bytes}
\approx
50\ \text{MB}.
$
On the same setup, this configuration achieves a throughput of approximately \(105{,}000\) tokens/s.

Assume now a global batch size of \(524{,}288\) tokens. Since each mini-batch contains $32 \times 1024 = 32{,}768$ tokens, the number of mini-batches that must be processed before performing a weight update is
$
\frac{524{,}288}{32{,}768}
=
16\ \text{mini-batches}
$.
Therefore, a complete training step consists of processing approximately 16 mini-batches. The gradients generated during the BW pass of each mini-batch are accumulated, and only after all 16 mini-batches have been processed, the model parameters are updated.
\end{example} 
The reason why these experiments were conducted using GPT-2 is that it is small enough to perform a training step on a single L40S accelerator. Models with a parameter count comparable to GPT-3 175B require substantially more memory than is available on a single GPU, making distributed training a necessity. This challenge is discussed in the next section.

\section{Distributed training: where traffic originates}
\label{ditributed_training}
So far, we have analyzed LLM training in a single device setting, following how the input tensor \(\mathbf{X}\) is processed through the model. In this section, we extend this view to distributed training. We first discuss why parallelization is required at the scale of large LLMs, and then analyze how different parallelization strategies partition the training workload across accelerators, generating the communication patterns and network traffic studied throughout this tutorial.

\subsection{The need for distributed training}
As shown in Example~\ref{single_gpu_mini_batch}, our single GPU setup is able to process approximately 105{,}000 tokens/s when training GPT-2. Assuming a training dataset containing $3 \times 10^9$ tokens, the total training time would be approximately
$
\frac{3 \times 10^9}{105{,}000}
\approx
28{,}500 \ \mathrm{seconds},
$
corresponding to roughly 8 hours of training on a single accelerator.
To understand why distributed training becomes necessary, let us progressively scale this example. GPT-2 contains approximately 124-128 million parameters, whereas larger models such as GPT-3 XL already contain about 1.3 billion parameters~\cite{brown2020language}, more than 10$\times$ larger. As a simplified approximation, let us assume that the achievable throughput decreases proportionally with model size. Under this assumption, the processing throughput would decrease from approximately \(10^5\) to \(10^4\) tokens/s. Now consider training such a model on a dataset containing \(10^{12}\) tokens. The required training time would become
$
\frac{10^{12}}{10^4} = 10^8 \ \mathrm{seconds},
$
which corresponds to more than 3 years of continuous training on a single GPU. Although highly simplified and based on several assumptions, this example highlights a fundamental challenge of modern LLM training: as model size and dataset size increase, training on a single accelerator rapidly becomes computationally impractical. Computation time, however, is only part of the problem: memory capacity quickly becomes an even more severe limitation. A simple estimate of the memory required to store a model can be obtained by multiplying the number of parameters by the storage size of each parameter. For example, assuming FP16 precision (2 bytes per parameter), a model containing 100 billion parameters requires approximately 200~GB of memory simply to store the parameters. During training, the memory requirements increase substantially due to the need to store gradients, optimizer states, and intermediate activations. In practice, the total memory footprint is often $4\times$ to $10\times$ larger than the model size alone. For instance, the activation memory required during GPT-3 training exceeds the memory occupied by the model parameters by more than 5$\times$~\cite{narayanan2021efficient}.
Consequently, training modern LLMs on a single accelerator becomes infeasible for two main reasons:
\begin{enumerate}
\item \textbf{Memory limitations: } the total memory required during training, including model parameters, activations, gradients, and optimizer states, exceeds the capacity of a single device. While current high end accelerators provide tens or, in some cases, hundreds of gigabytes of memory (e.g., NVIDIA H100~\cite{nvidia_h100}), these capacities remain insufficient for training the largest models.
\item \textbf{Computational time: } the amount of computation required to train models containing hundreds of billions or trillions of parameters, on datasets composed of trillions of tokens, would result in training times measured in years on a single accelerator.
\end{enumerate}
Despite continuous efforts by hardware vendors to increase the memory capacity available per accelerator, memory growth has not kept pace with the rapid scaling of LLMs. As reported in~\cite{hanindhito2024bandwidth}, model sizes have grown faster than the memory capacity of individual accelerators, creating an increasingly wide gap. Consequently, modern LLM training relies on distributed techniques that partition computation and memory across multiple accelerators.
\subsection{Parallelization Techniques and Communication Primitives}
The need to overcome the memory and computational limitations has led to the development of several distributed training strategies. These techniques differ in how the training workload is partitioned across multiple accelerators and, consequently, in the communication patterns they generate. Since network traffic is ultimately a consequence of how data, computations, and model parameters are distributed, understanding these parallelization strategies is essential for estimating the volume of data exchanged during training. In this section, we introduce the most widely adopted parallelization techniques and provide the background needed to derive their communication volumes. These techniques can be distinguished by the component of the training workload that they distribute across accelerators. The main strategies considered in this tutorial are:

\begin{itemize}
\item \textbf{Data Parallelism (DP):} the full model is replicated across multiple devices, and each device processes a different portion of the input data. The results are then combined to update a shared set of model parameters~\cite{rasley2020deepspeed, rajbhandari2020zero, zhao2023pytorch}.
\item \textbf{Pipeline Parallelism (PP):} the model is divided into sequential stages, each assigned to a different device, enabling different parts of the model to be executed concurrently on different portions of the input~\cite{huang2019gpipe, harlap2018pipedream, li2021terapipe, qi2024zero}.
\item \textbf{Tensor Parallelism (TP):} the computations within individual layers are split across multiple devices, allowing large model components to be distributed and processed in parallel~\cite{shoeybi2019megatron}. Sequence Parallelism (SP) extends TP by partitioning the input sequence across devices~\cite{korthikanti2023reducing} and is discussed in~\ref{app:dense_refinements}.
\item \textbf{Hybrid Parallelism:} DP, TP, SP, and PP are in practice often combined to enable efficient training of large scale LLMs~\cite{narayanan2021efficient, rasley2020deepspeed}.
\end{itemize} 
To establish a clear baseline, we consider simplified implementations of these strategies.~\ref{app:dense_refinements} discusses common optimizations and how different implementations alter the resulting traffic patterns. Duan et al.~\cite{duan2026efficient} and Rostam et al.~\cite{rostam2024achieving} review the distribution and optimization of LLM workloads across large scale systems, covering parallelization strategies, associated challenges, performance considerations, and the physical data center infrastructure (hardware connections, interconnects) used for training. MoE models introduce Expert Parallelism (EP)~\cite{shazeer2017outrageously, fedus2022switch}, which we discuss in~\ref{app:moe}.

Before examining the parallelization strategies, we introduce the required communication terminology. Distributed training uses \emph{Collective Communication Operations} (CCOs) to aggregate or redistribute tensors across accelerators, typically through \emph{Collective Communication Libraries} (CCLs). Their algorithms and implementations are discussed in Section~\ref{CCO_implementation}. The main collective considered here is \emph{AllReduce}, which aggregates a tensor across a group of accelerators and returns the result to every participant. In this section, we treat it as a logical synchronization operation and focus on the payload, participants, and invocation frequency.

We now analyze the main parallelization strategies.
Unless stated otherwise, the numerical examples in this section use the GPT-2-like configuration introduced previously, derived from the nanoGPT
implementation~\cite{karpathy_build_nanogpt_train_gpt2}. The model has
\(P\approx128\)M parameters, \(L=12\) Transformer layers, hidden dimension
\(d=768\), and sequence length \(T=1024\). All assumptions used in the examples are based on the performance characteristics of an NVIDIA L40s GPU equipped with \(48\,\mathrm{GB}\) of memory. The reported execution times were obtained from experiments conducted on the same hardware. Communicated activations and gradients use FP16/BF16 precision, such that \(s_{\mathrm{act}}=s_{\mathrm{grad}}=2\) bytes per value.
\subsubsection{Data Parallelism (DP)}
\begin{figure}[t]
    \centering
    \includegraphics[
    width=\linewidth,
    trim={0 0 0 0},
    clip
    ]{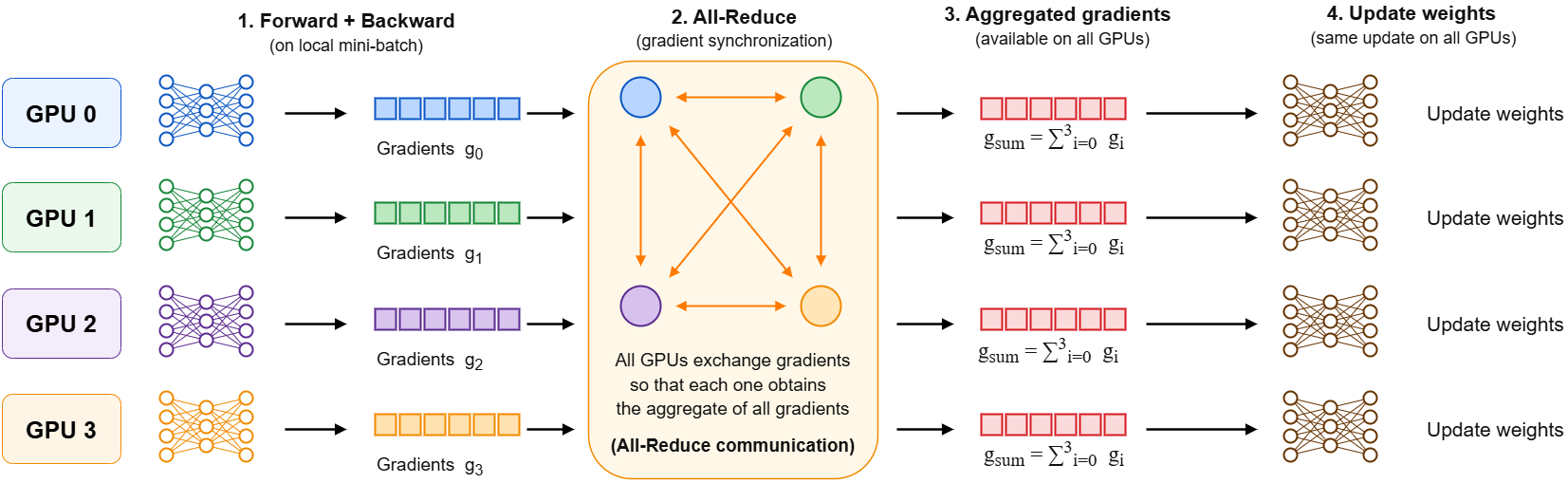}
    \caption{DP workflow. Each GPU maintains a replica of the model and
    processes a different mini-batch locally. After the FW and BW
    passes, the locally computed gradients are exchanged and aggregated across
    all GPUs through an AllReduce CCO.}

    \label{fig:DP_communication}
\end{figure}
To simplify the discussion, we begin by assuming that the entire model can fit within the memory of a single accelerator. The idea behind DP is to train several identical copies of the same model in parallel. Each accelerator stores a full replica of the model, but processes a different portion of the training data. The FW and BW passes are performed independently on each accelerator, producing local gradients from the data processed on that device. When a parameter update is required, these gradients are synchronized across all accelerators in the DP group, typically through an \emph{AllReduce} CCO. After synchronization, each accelerator applies the same weight update, ensuring that all model replicas remain identical. In this way, DP increases the amount of training data processed in parallel while keeping all copies of the model synchronized. Fig.~\ref{fig:DP_communication} illustrates this execution flow, highlighting the local computation performed on each GPU and the subsequent gradient synchronization phase. From a communication perspective, the total number of gradient values is equal to the number of model parameters. Consequently, whenever gradients are synchronized, each accelerator participates in the synchronization of a gradient tensor whose size is approximately equal to the model size. Given \(P\) and \(s_{\mathrm{grad}}\), the logical gradient tensor synchronized by each DP worker has size $M_{\mathrm{DP}}$ defined in~\eqref{eq:gradient_payload}.
For FP16/BF16 gradient communication, \(s_{\mathrm{grad}}=2\) bytes, and thus $M_{\mathrm{DP}} = P \times 2~\mathrm{bytes}$. In standard DP, gradient synchronization is usually implemented through an
AllReduce operation.
The following example applies the reference configuration to eight
DP workers.
\begin{example}[Gradient Synchronization in DP]
\label{DP_batch}
Using the reference configuration, consider DP across
\(p_{\mathrm{DP}}=8\) GPUs. Each GPU stores a full model replica and processes a local mini-batch of \(B=32\) sequences. The target global batch size is \(B_{\mathrm{global}}=524{,}288\) tokens. The number of gradient accumulation steps per GPU is
$
N_{\mathrm{acc}}
=
\frac{B_{\mathrm{global}}}
     {B T p_{\mathrm{DP}}}
=
\frac{524{,}288}
     {32\cdot1024\cdot8}
=
2.
$
Thus, before each optimizer update, every GPU processes two local mini-batches,
corresponding to
$
2 \cdot 32 \cdot 1024
=
65{,}536
\text{ tokens per GPU}.
$
Across all eight GPUs, this gives the target global batch size:
$
8 \cdot 65{,}536
=
524{,}288
\text{ tokens}.
$
After local gradient accumulation, the DP workers synchronize their gradients.
The logical gradient tensor synchronized by each worker has size
$
M_{\mathrm{DP}}
=
P s_{\mathrm{grad}}
=
256 \times 10^6
\text{ bytes}
\approx 256\,\mathrm{MB}.
$
Therefore, each GPU participates in one
AllReduce over an approximately \(256\,\mathrm{MB}\) logical payload per
optimizer update. The communication volume generated by the collective
implementation is derived in Section~\ref{CCO_implementation}.
\end{example}

This example highlights the main communication characteristic of DP: communication is relatively infrequent, since it occurs once per optimizer update, but the synchronized object is proportional to the model size. Therefore, DP becomes network-intensive when the model is large, when many DP workers participate in the synchronization, or when optimizer updates occur frequently.

\subsubsection{Pipeline Parallelism (PP)}
\begin{figure}[t]
    \centering
    \includegraphics[
    width=\linewidth,
    trim={10 0 10 0},
    clip
    ]{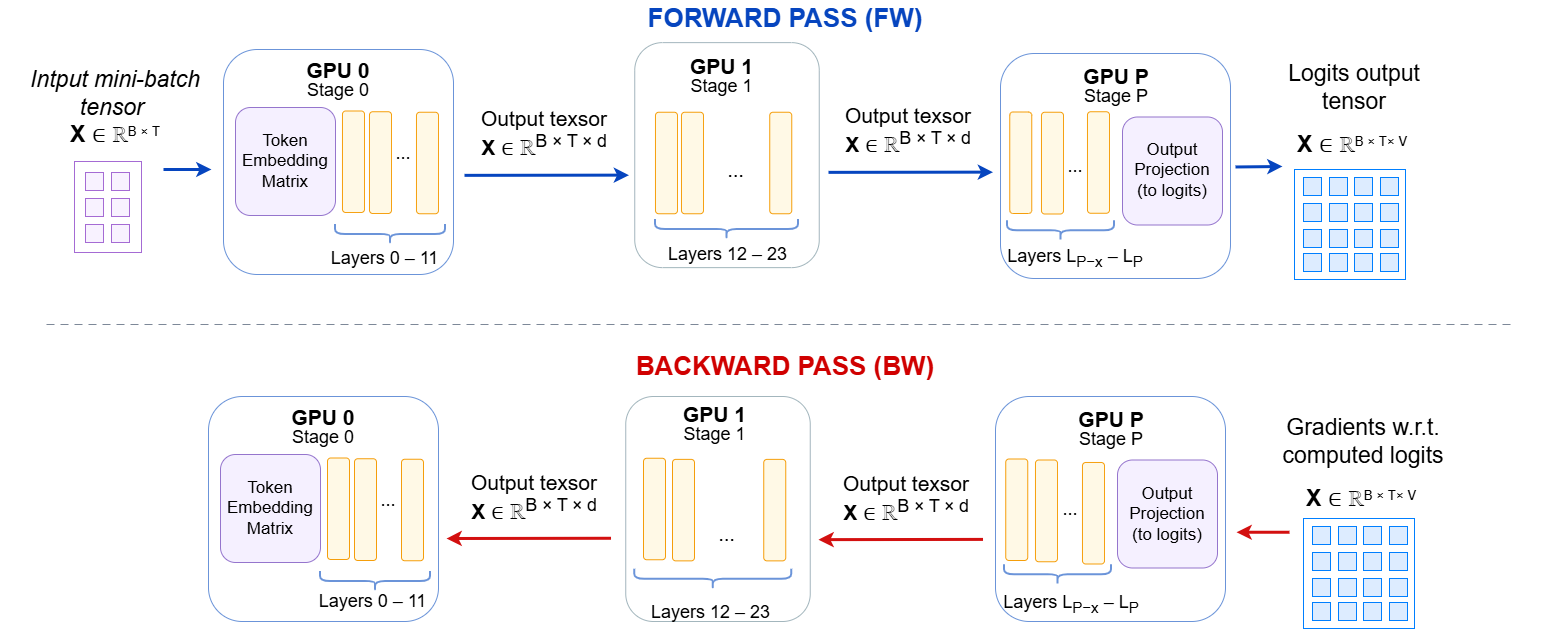}
    \caption{PP workflow. Transformer layers are partitioned across multiple
    GPUs, while the activation tensor $\mathbf{X}$ is exchanged between
    adjacent pipeline stages during the FW and BW passes through
    point-to-point communication.}
    \label{fig:PP_communication}
\end{figure}

PP represents a second widely adopted parallelization strategy for LLM training. In this case, we start from the assumption that the model is too large to fit within the memory of a single accelerator. The idea behind PP is to split the Transformer stack across multiple accelerators, so that each device stores and executes only a consecutive subset of layers. For example, a model with 96 Transformer layers, such as GPT-3~\cite{brown2020language}, can be distributed across 8 GPUs, with each GPU storing 12 layers instead of the full model.
During the FW pass, the input is first processed by the accelerator hosting the initial layers of the model. The resulting activation tensor is then sent to the next accelerator, which processes the following layers. This process continues stage by stage until the final accelerator produces the model output. During the BW pass, gradients follow the same pipeline in the opposite direction, moving from the last stage back toward the first.
From a communication perspective, PP differs from DP because data is not synchronized across all accelerators. Instead, communication mainly consists of P2P exchanges between adjacent pipeline stages. For each processed input, the activation tensor is transmitted FW across a pipeline boundary, while the corresponding activation gradient tensor is transmitted BW across the same boundary. These activation and activation gradient exchanges therefore represent the primary source of network traffic in PP training. Fig.~\ref{fig:PP_communication} illustrates the execution flow of PP, highlighting how the activation tensor propagates across adjacent pipeline stages during the FW and BW passes.
For a mini-batch of size \(B\), sequence length \(T\), hidden dimension \(d\),
and activation precision \(s_{\mathrm{act}}\) bytes, the activation tensor
exchanged between two adjacent pipeline stages has the shape in ~\eqref{eq:activation_tensor_shape} and the payload $M_{\mathrm{PP}}$ defined in~\eqref{eq:activation_payload}.
Since the activation tensor is transmitted once during the FW pass and its corresponding gradient once during the BW pass, the communication volume across one pipeline boundary is $C_{\mathrm{PP,boundary}} \approx 2M_{\mathrm{PP}} = 2S_{\mathrm{act}}$ where $S_{\mathrm{act}}$ is the activation payload defined in~\eqref{eq:activation_payload}. This volume is local to each pipeline boundary: unlike DP, the tensor is not synchronized across all accelerators, but exchanged only between neighboring pipeline stages. The following example instantiates this calculation for a GPT-2-like model partitioned across eight pipeline stages. The example deliberately uses a naive non-overlapped pipeline execution in order to expose the size and timing of the activation transfers before introducing bubble-reducing schedules.

\begin{example}[Activation Transfers in PP]
\label{PP_batch}

Using the reference configuration, consider \(p_{\mathrm{PP}}=8\) pipeline stages processing a mini-batch of \(B=32\) sequences. We consider a naive pipeline partition where the \(12\) Transformer layers are distributed across the \(8\) GPUs as $[1,1,2,2,2,2,1,1]$.
Thus, the first two and last two stages contain one Transformer layer each, while the four middle stages contain two layers each. This partition is deliberately unbalanced to account for the additional input and output vocabulary layers assigned to the first and last stages, which can create both computational and memory imbalances
~\cite{tsung2025balancing}. Assume the following per-layer execution times:
$
t_{\mathrm{FW}}^{\mathrm{layer}} \approx 20\,\mathrm{ms}$, and $
t_{\mathrm{BW}}^{\mathrm{layer}} \approx 40\,\mathrm{ms}.
$
The FW pass execution times of the eight stages are therefore
$
[20,20,40,40,40,40,20,20]\,\mathrm{ms},
$
which gives a total naive FW pipeline latency of
$
20+20+40+40+40+40+20+20 = 240\,\mathrm{ms}.
$
Similarly, the naive BW pipeline latency is approximately $2 \times 240 = 480\,\mathrm{ms}.$ At each pipeline boundary, adjacent stages exchange the activation
tensor defined in~\eqref{eq:activation_tensor_shape}. Its logical
payload is
$
M_{\mathrm{PP}}
=
S_{\mathrm{act}}
=
32 \cdot 1024 \cdot 768 \cdot 2
\approx
50\,\mathrm{MB}
$
where \(S_{\mathrm{act}}\) is defined in Eq.~\eqref{eq:activation_payload}. Each pipeline boundary therefore carries approximately \(50\,\mathrm{MB}\) during the FW pass and the same amount of activation-gradient data during the BW pass, for a total of approximately \(100\,\mathrm{MB}\) per mini-batch. In this naive non-overlapped execution, a full-mini-batch activation transfer therefore occurs across each pipeline boundary once during the approximately \(240\,\mathrm{ms}\) FW traversal, and the corresponding
activation-gradient transfer occurs once during the approximately
\(480\,\mathrm{ms}\) BW traversal.
\end{example}
This example highlights the main communication characteristic of PP: communication is activation-sized and local, as tensors are exchanged only between adjacent pipeline stages. However, these transfers lie on the critical path, and their dependencies may leave stages idle, creating \emph{pipeline bubbles}. Practical PP implementations divide each mini-batch into \(m\) \emph{micro-batches} of size \(b_{\mu}\), such that \(B=m b_{\mu}\). Each micro-batch is propagated independently through the pipeline, and the tensor exchanged between adjacent stages is
$
\mathbf{X}_i\in\mathbb{R}^{b_{\mu}\times T\times d},
i\in\{1,\ldots,m\}.
$
Assuming equal precision for activations and their gradients, the communication volume per micro-batch and boundary is
$
2b_{\mu}Td\,s_{\mathrm{act}}.
$
Across all micro-batches, the total remains
$
\sum_{i=1}^{m}2b_{\mu}Td\,s_{\mathrm{act}}
=2S_{\mathrm{act}}.
$
Micro-batching therefore changes the granularity and timing of transfers rather than their total volume. Scheduling techniques that exploit this decomposition are introduced in~\ref{app:dense_refinements}.
\begin{figure}[t]
    \centering
    \includegraphics[
    width=\linewidth,
    trim={0 0 0 0},
    clip
    ]{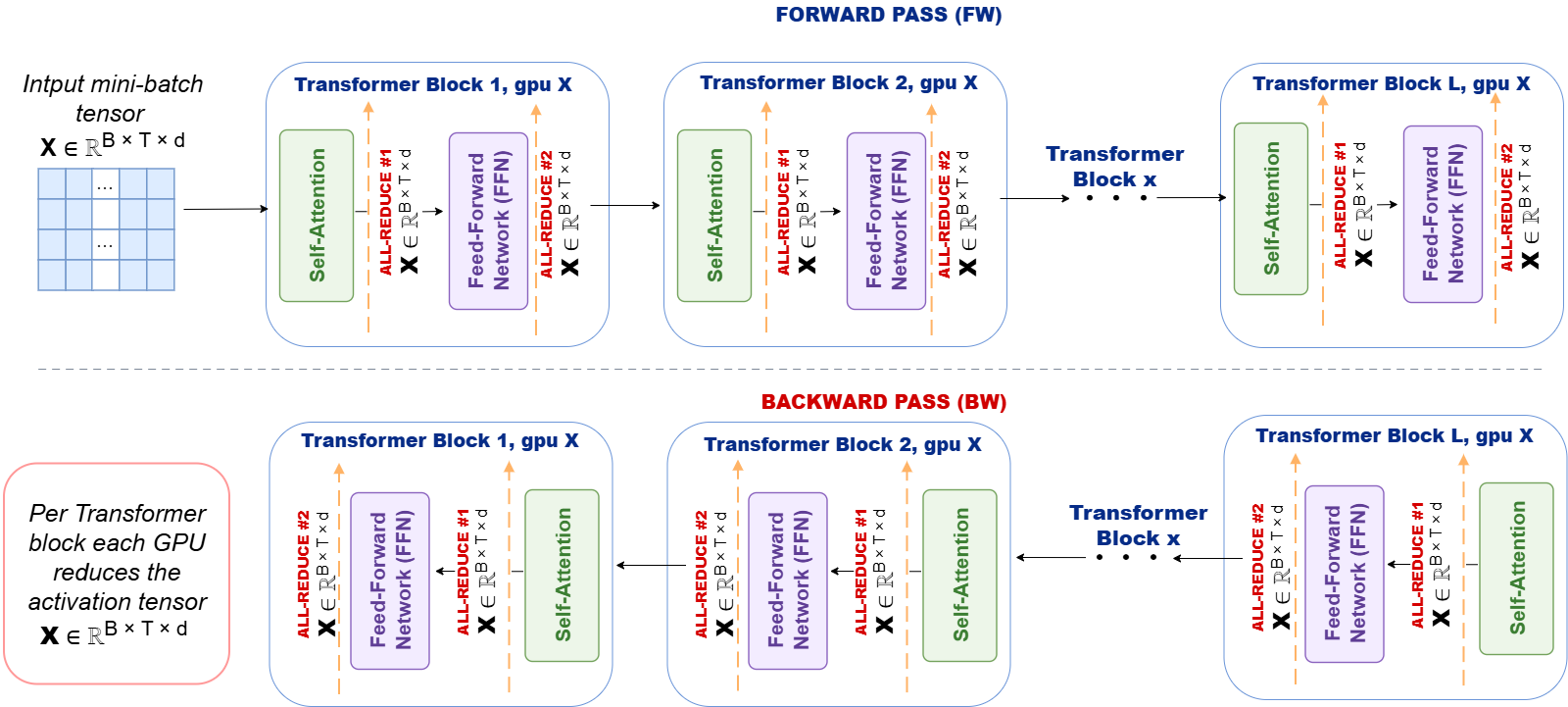}
    \caption{TP workflow: Transformer blocks span multiple GPUs, with CCOs exchanging $\mathbf{X}$ during FW and BW.}
    \label{fig:TP_communication}
\end{figure}
\subsubsection{Tensor Parallelism (TP)}
As in the case of pipeline parallelism, we assume that the model is too large to fit within the memory of a single accelerator. Unlike PP, however, TP partitions the computation occurring inside individual Transformer layers across multiple GPUs. In particular, the matrix multiplications performed in the self-attention and FFN blocks are distributed across several accelerators, each of which computes only a portion of the operation. The detailed partitioning schemes are described in the original Megatron-LM work~\cite{shoeybi2019megatron}, while an intuitive pedagogical walkthrough of these operations is also available online~\cite{won_tensor_parallelism}. In this tutorial, we focus primarily on the resulting communication patterns. Compared to DP and PP, TP introduces significantly more frequent communication, since synchronization occurs inside every Transformer layer. As discussed in~\cite{shoeybi2019megatron}, tensor-parallel execution typically requires two AllReduce CCOs per Transformer layer during the FW pass and two AllReduce CCOs during the BW pass. Fig.~\ref{fig:TP_communication} illustrates the execution flow of TP, highlighting the CCOs performed inside each Transformer block during the FW and BW passes.
From a communication perspective, the tensor exchanged by TP is associated with
the activation tensor processed by the Transformer layer. The activation tensor involved in TP communication has the shape given in~\eqref{eq:activation_tensor_shape} and the payload defined in~\eqref{eq:activation_payload}. Since TP typically requires two AllReduce operations in the FW pass and two
AllReduce operations in the BW pass for each Transformer layer, the
logical amount of activation data synchronized per layer is approximately $4M_{\mathrm{TP}}$. The following example instantiates this calculation for a GPT-2-like model
trained with TP across eight GPUs. 

\begin{example}[AllReduce Payload in TP]
\label{TP_batch}
Using the reference configuration, consider TP across
\(p_{\mathrm{TP}}=8\) GPUs processing a mini-batch of \(B=32\) sequences. The communicated activation tensor has the shape defined in~\eqref{eq:activation_tensor_shape} and the payload given in~\eqref{eq:activation_payload}. Substituting the above values gives
$
M_{\mathrm{TP}}
=
S_{\mathrm{act}}
=
32 \cdot 1024 \cdot 768 \cdot 2
\approx
50\,\mathrm{MB}.
$
Assume that the execution of a single Transformer layer is distributed across eight GPUs and requires approximately
$
t_{\mathrm{FW}} \approx 3\,\mathrm{ms}$, and $
t_{\mathrm{BW}} \approx 6\,\mathrm{ms}.
$
In the baseline non optimized TP implementation, each Transformer layer performs two AllReduce operations during the FW pass and two during the BW pass, each over a logical payload of approximately \(50.3\,\mathrm{MB}\). Under the assumed
execution times, this corresponds to one AllReduce every \(1.5\,\mathrm{ms}\) during the FW pass and every \(3\,\mathrm{ms}\) during the BW pass.
\end{example}

This example highlights the main communication characteristic of TP: the communicated tensor is activation-sized, as in PP, but the communication is much more frequent because it occurs multiple times inside every Transformer layer. Therefore, TP is sensitive not only to bandwidth, but also to communication latency and synchronization overhead. The main limitation of TP is that it is highly communication intensive, requiring synchronization operations much more frequently than DP or PP. For this reason, TP strongly depends on extremely high-bandwidth and low-latency interconnects, and becomes significantly less efficient when accelerators are distributed across different nodes without ultra-fast communication links. Moreover, compared to DP and PP, TP is generally more complex to implement and adapt to evolving Transformer architectures.
\begin{table}[t]
\centering
\caption{Distributed Training Communication Patterns}
\label{tab:parallelism_summary}
\begin{tabularx}{\columnwidth}{lX}
\toprule
\textbf{Technique} & \textbf{Communication Pattern} \\
\midrule

DP &
Model sized gradient synchronization with payload
\(S_{\mathrm{grad}}\) defined in~\eqref{eq:gradient_payload} across data parallel replicas, typically through AllReduce. \\

PP &
Activation and activation-gradient tensors exchanged through P2P communication between adjacent pipeline stages with activation tensor
shape defined in~\eqref{eq:activation_tensor_shape}. \\

TP &
Multiple CCOs inside each Transformer layer during both the FW and BW passes. The exchanged tensor has the shape defined in~\eqref{eq:activation_tensor_shape}. SP, a TP optimization, follows the same general structure as TP, but with different tensor partitioning and execution scheduling. \\

Hybrid &
Combination, placement, and overlap of the communication patterns generated by DP, PP, TP and SP. \\

\bottomrule
\end{tabularx}
\end{table}
\subsubsection{Hybrid Parallelism and communication summary}
In practical large-scale LLM training, parallelization strategies are rarely used in isolation. Instead, modern systems combine multiple dimensions of parallelism in order to jointly address model memory constraints, computational scalability, and communication efficiency. This combination of multiple strategies is commonly referred to as \emph{hybrid parallelism}. In this setting, \emph{3D parallelism} usually denotes the joint use of DP, TP, and PP, while \emph{4D parallelism} extends this design by introducing an additional dimension, often SP or another context-related partitioning strategy ~\cite{narayanan2021efficient,korthikanti2023reducing,singh20234d}. Hybrid parallelism has become the dominant paradigm for large-scale LLM training. It is employed by many widely adopted systems and frameworks, including Megatron-LM~\cite{narayanan2021efficient}, DeepSpeed ~\cite{rasley2020deepspeed}, Mesh-TensorFlow~\cite{shazeer2018mesh}, and production-scale training systems such as PaLM~\cite{chowdhery2023palm}. Although these systems differ in their implementation strategies, they all rely on combining multiple forms of parallelism across different groups of accelerators. More comprehensive overviews of distributed training strategies and optimization techniques are provided in~\cite{duan2026efficient,zeng2025distributed}. From a communication perspective, hybrid parallelism should not be viewed as a completely new communication mechanism. Instead, it combines and overlaps the traffic patterns generated by its constituent strategies. Communication partitioning and hierarchical scheduling can further improve the overlap between these exchanges and model computation~\cite{chen2024centauri}. DP introduces model-sized gradient synchronization across model replicas, PP generates point-to-point exchanges of activation tensors between adjacent pipeline stages, and TP requires frequent collective communication inside Transformer layers. Consequently, the simplified formulations introduced in the previous subsections remain the fundamental building blocks for characterizing network traffic also in hybrid training configurations as well. Table~\ref{tab:parallelism_summary} summarizes these communication patterns and

\begin{figure}[t]
    \centering
    \includegraphics[
    width=\linewidth,
    trim={0 0 0 0},
    clip
    ]{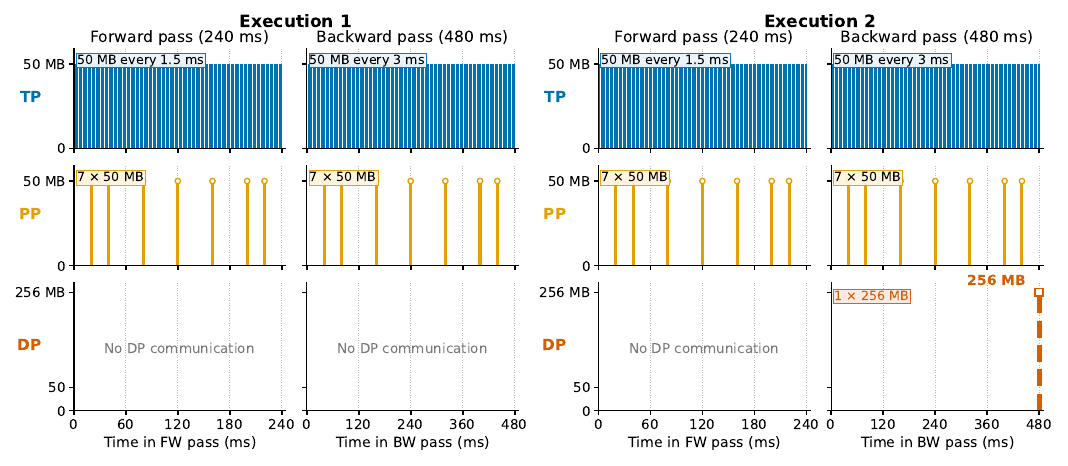}
    \caption{Communication schedule for a full batch execution, composed of two consecutive executions and showing TP, PP, and DP transfers during the forward and backward passes.}
    \label{fig:commSchema}
\end{figure}

Fig.~\ref{fig:commSchema} shows how the traffic patterns from the previous examples combine in a simple 3D configuration. TP generates frequent \(50\,\mathrm{MB}\) AllReduce operations, PP exchanges \(50\,\mathrm{MB}\) activation tensors at stage boundaries via simple P2P operations, and DP performs one \(256\,\mathrm{MB}\) AllReduce after two gradient accumulation steps. The figure makes it clear that the three parallelization strategies do not create a new traffic pattern when combined. Instead, the traffic generated by DP, PP, and TP coexists within the same training execution, with each strategy contributing its own communication pattern. The timeline is illustrative and compares communication frequency and payload size; it does not represent a measured end-to-end trace of a fully integrated hybrid training configuration. This traffic pattern is consistent with several studies of network behavior during LLM training~\cite{hanindhito2024bandwidth, anthony2024demystifying, gangidi2024rdma, qian2024alibaba, wang2024rail}. Gangidi et al.~\cite{gangidi2024rdma} describe it as predictable and repetitive, with few connections and millisecond scale on off bursts. During each active burst, large tensors are transferred as quickly as the network permits, so the instantaneous data rate can approach the available link bandwidth. However, the average rate over the full training step is lower because communication bursts alternate with periods of computation in which little or no data is transferred. 

\begin{figure}[t]
    \centering
    \includegraphics[
    width=\linewidth,
    trim={10 0 10 0},
    clip
    ]{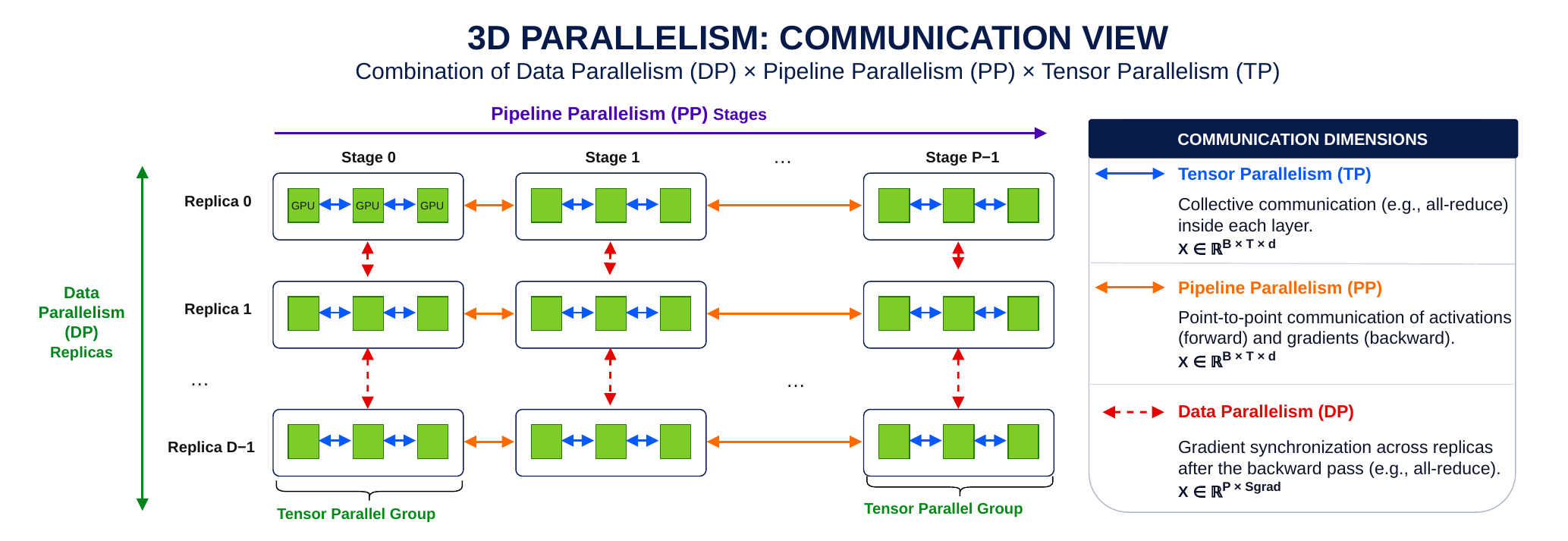}
    \caption{Example of 3D hybrid parallelism combining DP, PP, and TP. GPUs are organized in a 3D grid; different parallel dimensions generate different communication patterns.}

    \label{fig:parallelization_communication_summary}
\end{figure}

Selecting a specific hybrid parallelization strategy depends on several factors, including model size, sequence length, accelerator memory capacity, and the characteristics of the underlying network topology and infrastructure discussed in Section~\ref{net_top_infra}. The input length distribution also matters, since variable length documents can create uneven computation and communication loads across pipeline groups~\cite{wang2025wlbllm}. For this reason, several works have explored automatic approaches for deriving efficient hybrid configurations, including Alpa~\cite{zheng2022alpa}, FlexFlow~\cite{jia2019beyond}, GSPMD~\cite{xu2021gspmd}, Unity~\cite{unger2022unity}, Galvatron~\cite{miao2022galvatron}, and
Aceso~\cite{liu2024aceso}. Nevertheless, determining the optimal hybrid parallelization strategy for a given training scenario remains an open research problem. Despite this, a commonly adopted design pattern is to map TP groups onto accelerators connected through extremely high-bandwidth and low-latency links, typically within the same node. Conversely, PP and DP are generally more suitable for communication across nodes, where network performance is comparatively lower~\cite{jiang2024megascale,chu2025scaling}. Fig.~\ref{fig:parallelization_communication_summary} summarizes this placement intuition and the resulting communication patterns.

\section{CCOs: from tensor to communication payloads}
\label{CCO_implementation}
In the previous section, we identified the tensor payloads that are collectively communicated under different distributed training strategies. In the Life of a Token framework, this marks the transition from tensors as computational objects to tensors as communication objects. This section examines how such tensors are processed by CCOs and how the implementation of these operations affects the resulting network traffic. This analysis matters because a tensor size alone is not sufficient to determine communication volume: the same CCO can be implemented through different algorithms, each inducing a different communication schedule, number of transfers, and traffic distribution across links. To illustrate how collective algorithms move, combine, and redistribute data, we focus on \emph{AllReduce}, one of the most widely used CCOs in our scenario. As discussed in the previous section, it plays a central role in gradient synchronization during large-scale LLM training. It is also a useful tutorial example because, depending on its implementation, it can be decomposed into simpler operations, such as \emph{ReduceScatter}, \emph{AllGather}, \emph{Reduce}, and \emph{Broadcast}, allowing us to introduce multiple CCOs through a single representative operation.
This section is therefore not intended as a complete survey of CCOs or CCLs. Instead, it provides a conceptual bridge for characterizing how the tensor payloads generated during LLM training translate into the network traffic induced by their collective communication. A broader catalogue of collective operations is available in the NCCL documentation~\cite{nvidia_nccl_collectives}.

\subsection{AllReduce implementation strategies}
We now examine how different implementations of \emph{AllReduce} translate the same logical tensor payload into different communication volumes. The key question is how the reduction is scheduled, since the schedule determines the number of transfers and the resulting traffic pattern. To make this effect explicit, we compare three representative implementations: a naive AllReduce, a ring-based AllReduce, and a tree-based AllReduce. The naive version provides a simple baseline, while the ring and tree variants illustrate two widely used design principles for organizing collective communication.
\subsubsection{Naive AllReduce} A simple but inefficient implementation of AllReduce consists of having each GPU broadcast its local tensor to all the GPUs participating in the operation. After receiving the tensors from every participant, each accelerator locally computes the final reduction. Although intuitive, this method generates a very large amount of network traffic and scales poorly as the number of accelerators increases. In this naive implementation, each of the $p_{\mathrm{cc}}$ GPUs sends a message of size $M$ to all other $p_{\mathrm{cc}}-1$ GPUs. Therefore, the traffic generated by each GPU is:
$
C_{\mathrm{dev}}^{\mathrm{naive}}
\left(M,p_{\mathrm{cc}}\right)
=
(p_{\mathrm{cc}}-1)M.
$
The total traffic across the system is:
$
C_{\mathrm{total}}^{\mathrm{naive}} =
p_{\mathrm{cc}}C_{\mathrm{dev}}^{\mathrm{naive}}=
p_{\mathrm{cc}}(p_{\mathrm{cc}}-1)M.
$
\subsubsection{Ring AllReduce}
Optimized AllReduce implementations avoid the naive broadcast based behavior. The key idea is to arrange the $p_{\mathrm{cc}}$ participating GPUs in a logical ring and split the exchanged tensor $M$ into $p_{\mathrm{cc}}$ equally sized chunks. Instead of sending the entire tensor to every other GPU, each accelerator exchanges one chunk at a time with its neighbors in the ring. \textit{Ring AllReduce} is typically implemented in two phases. The first phase is the \emph{ReduceScatter} phase: tensor chunks circulate around the ring, and partial reductions are accumulated until each GPU holds one fully reduced chunk. The second phase is \emph{AllGather}: the reduced chunks circulate again so that, at the end of the operation, every GPU reconstructs the complete reduced tensor.  This implementation is bandwidth efficient for large tensors because each GPU communicates only with its two ring neighbors and transfers data in a regular chunked pattern. As a result, communication can be pipelined across the ring, and each P2P exchange can use the available bandwidth of the corresponding link, assuming no external contention. This makes ring AllReduce a widely adopted implementation. With this approach, the amount of data sent by each GPU is approximately:
$
C_{\mathrm{GPU}}^{\mathrm{ring}} =
2 \cdot \frac{p_{\mathrm{cc}} - 1}{p_{\mathrm{cc}}} \cdot M.
$
Consequently, the total amount of data transmitted across the system is:
$
C_{\mathrm{total}}^{\mathrm{ring}} =
p_{\mathrm{cc}} \cdot C_{\mathrm{GPU}}^{\mathrm{ring}}
=
2(p_{\mathrm{cc}} - 1)M.
$
Fig.~\ref{fig:ringAllReduce} summarizes the execution of ring AllReduce across four GPUs.

\subsubsection{Tree based AllReduce}
\begin{figure}[!t]
    \centering
    \subfloat[Example of Ring AllReduce execution across four GPUs
    \label{fig:ringAllReduce}]{
        \includegraphics[width=0.45\linewidth]{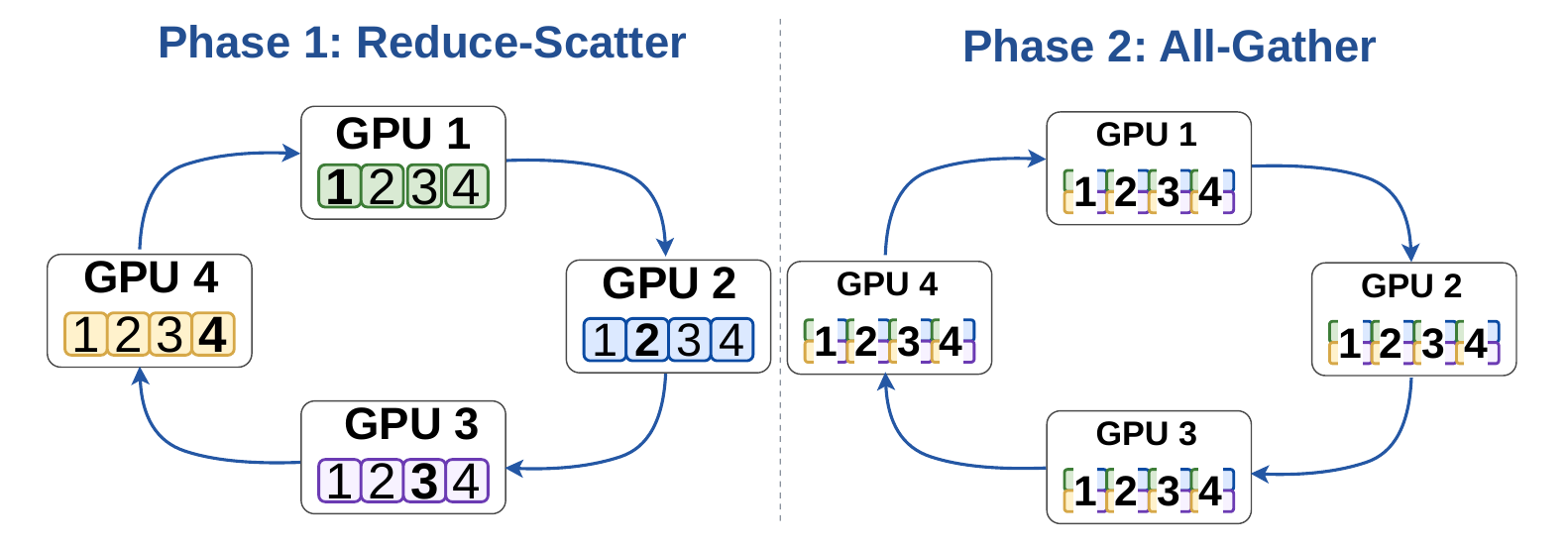}
    }%
    \hfill
    \subfloat[Example of Tree AllReduce execution across four GPUs
    \label{fig:treeAllReduce}]{
        \includegraphics[width=0.45\linewidth]{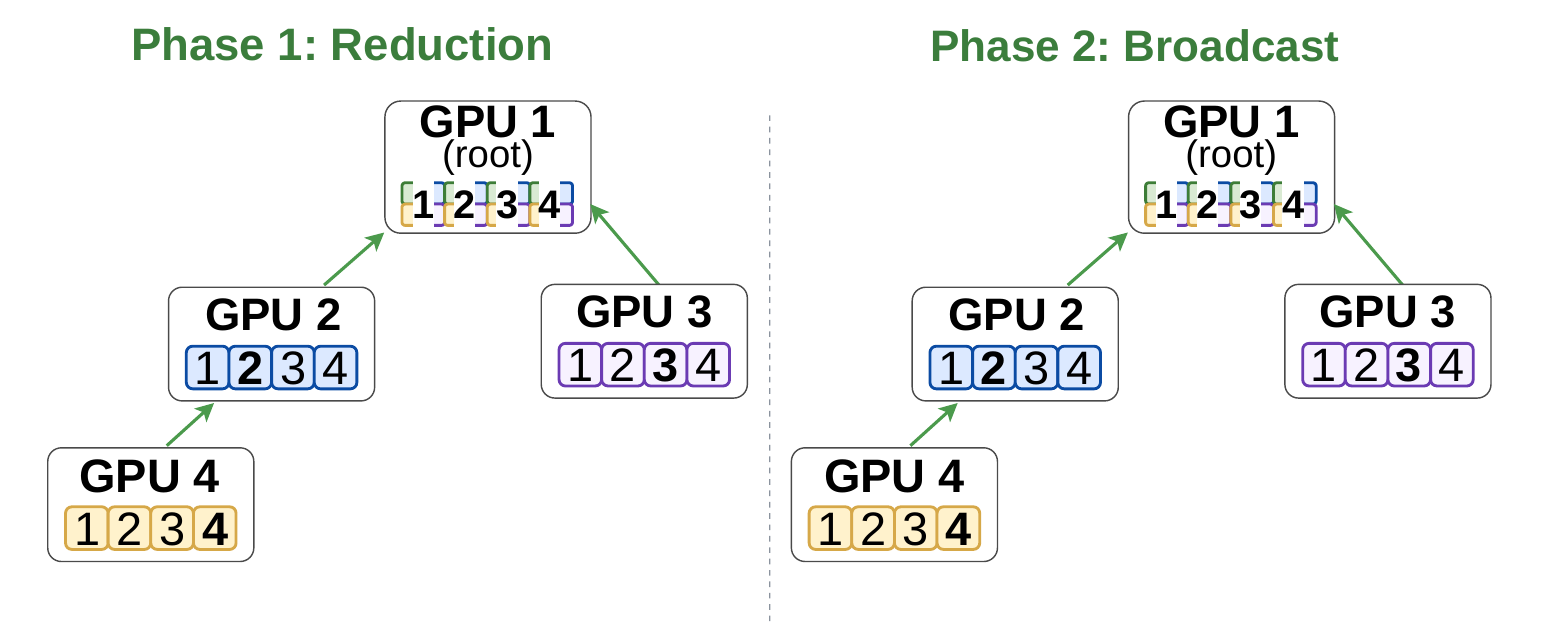}
    }
    \caption{Comparison of Ring and Tree AllReduce for four GPUs and their two communication phases.}

    \label{fig:CCO_comparison}
\end{figure}

Tree-based AllReduce provides an alternative implementation that is often better suited for smaller tensors or latency sensitive communication. Instead of arranging GPUs in a ring, the participating accelerators are organized into a logical tree. Communication then proceeds hierarchically along the tree structure. As in ring AllReduce, the operation can be interpreted as two phases. In the first phase, \emph{reduce}, data moves from the leaves of the tree toward the root. At each internal node, partial tensors received from children are combined with the local tensor, until the root obtains the fully reduced result. In the second phase, \emph{broadcast}, the reduced tensor is propagated from the root back down the tree so that every GPU receives the final result. 
For a payload of size $M$, each edge of the tree carries approximately $M$ bytes during the reduce phase and $M$ bytes during the broadcast phase. Since a tree with $p_{\mathrm{cc}}$ GPUs contains $p_{\mathrm{cc}}-1$ edges, the total amount of data transmitted across the system is:
$
C_{\mathrm{total}}^{\mathrm{tree}} =
2(p_{\mathrm{cc}}-1)M.
$
From the perspective of a single GPU, the amount of data sent depends on its position in the tree. 
Leaf nodes send data during the reduce phase and receive data during the broadcast phase, while internal nodes may send and receive data from multiple neighbors. Therefore, tree based AllReduce is often better characterized either by its total system traffic or by its number of communication steps. Averaged over all GPUs, the communication volume per GPU is: 
$
C_{\mathrm{GPU,avg}}^{\mathrm{tree}} =
\frac{C_{\mathrm{total}}^{\mathrm{tree}}}{p_{\mathrm{cc}}}
=
2 \frac{p_{\mathrm{cc}}-1}{p_{\mathrm{cc}}} M.
$
Assuming a balanced binary tree, the number of communication steps is approximately $2\log_2 {p_{\mathrm{cc}}}$.
Tree-based AllReduce is especially useful when the message size $M$ is small or when latency dominates communication time. In this regime, the number of sequential communication steps becomes more important than the total amount of data transferred. A balanced tree completes the reduce and broadcast phases in approximately $2\log_2 p_{\mathrm{cc}}$ steps, whereas ring AllReduce requires a number of steps that grows linearly with $p_{\mathrm{cc}}$. For this reason, tree-based algorithms can reduce synchronization latency for small messages, while ring-based algorithms are typically preferred for large tensors where bandwidth efficiency dominates. Fig.~\ref{fig:treeAllReduce} illustrates the execution of Tree AllReduce across four GPUs.
\subsection{Collective Communication Libraries}
Parallelization strategies express their required tensor exchanges through CCOs, which define the logical communication among accelerator groups without specifying its hardware implementation. CCLs provide this implementation. A CCL takes a logical CCO and maps it to an executable communication schedule across the available devices. In doing so, it selects or instantiates a collective algorithm, such as a ring, tree, or specific topology aware schemes, divides tensors into chunks, organizes the order of transfers, and coordinates the synchronization among ranks. In practice, the CCL decides the communication schedule followed by the data: which devices talk to each other, which transfers stay inside a node, and which ones cross the network between nodes. The underlying network then carries these transfers over the available interconnects and routes. This distinction is important for traffic analysis. The same tensor payload and the same CCO can lead to different communication patterns depending on how the operation is implemented. For example, an AllReduce based on a ring algorithm results in a sequence of pairwise exchanges, while a tree-based implementation organizes communication in a hierarchical manner. In addition, topology aware implementations can decide whether to keep communication within a node using faster local paths or to route data across nodes when needed, depending on the system layout. These choices affect how data is partitioned, scheduled, and exchanged across devices. Widely used CCLs include NVIDIA NCCL~\cite{nccl}, which is commonly used for GPU-based distributed training, as well as RCCL~\cite{rccl} for AMD GPUs, HCCL for Huawei accelerators, PyTorch Gloo~\cite{pytorch_gloo}, and Intel oneCCL~\cite{intel_oneccl}. Recent work has analyzed the behavior of these libraries and explored improved collective implementations through topology aware algorithm design, optimized scheduling, and more flexible communication backends~\cite{hu2025demystifying,weingram2023xccl,cai2021synthesizing,shah2023taccl,liu2024rethinking,cowan2023mscclang,shah2025msccl}. Low level CCL parameters can also be tuned automatically for the hardware configuration and workload, including the interference between communication and computation~\cite{xu2025autoccl}. In the context of this paper, CCLs are relevant because they form the bridge between payload-level estimates and the traffic eventually observed by the network. They determine how abstract communication primitives are decomposed into concrete transfers, making them essential for connecting LLM training tensors to link-level communication behavior.

\subsection{Communication Estimates in Practice}
\label{subsec:volume_to_measurements}
The communication volume generated by a CCO depends on both the tensor payload and the algorithm used to implement the operation. Although this tutorial uses AllReduce as a representative example, the same reasoning applies to other CCOs, whose implementations may follow different schedules and therefore generate different traffic patterns~\cite{weingram2023xccl, thakur2005optimization}.  Although Ring and Tree AllReduce move approximately the same total amount of data, they divide this transfer into a different number of communication rounds. Each round introduces a fixed startup cost, including operation launch, synchronization, and message-processing overhead, even when the message itself is very small. Ring AllReduce requires approximately \(2(p_{\mathrm{cc}}-1)\) sequential rounds, because both its ReduceScatter and AllGather phases proceed around the ring. For small tensors, the time required to transmit the data is short, so these repeated startup costs dominate the overall execution time. A tree reduces the number of sequential rounds to approximately \(\mathcal{O}(\log p_{\mathrm{cc}})\), allowing small messages to complete with fewer synchronization delays.
For large tensors, the data-transfer time becomes more important than the startup cost. Ring AllReduce divides the tensor into chunks and pipelines them around the ring, allowing all accelerators to send and receive data simultaneously. Its traffic is also distributed evenly across the participants, which helps sustain high link utilization and approach the available bandwidth. In a tree, communication is concentrated along the tree levels, and links or accelerators closer to the root may become bottlenecks. Tree-based implementations may therefore complete fewer rounds but achieve lower sustained bandwidth for large tensors. Consequently, Tree AllReduce is often preferred for small or latency-sensitive messages, whereas Ring AllReduce is usually better suited for large, bandwidth-dominated transfers~\cite{thakur2005optimization,weingram2023xccl}. Although payload size and collective algorithm determine the baseline communication demand, completion time can also be affected by stragglers and network congestion, particularly in shared environments
~\cite{warraich2025optireduce}.

\section{Network infrastructure: from payloads to bits on the wire}
\label{net_top_infra}
Network infrastructure and interconnect technologies form the fifth element of the proposed framework, linking payload-level communication requirements to the actual movement of data across the training system. The previous stages determine the size of the tensors to be exchanged, the logical communication patterns induced by the adopted parallelization strategy, and the communication volume shaped by CCO algorithms and their implementation inside CCLs. Once these logical exchanges are scheduled, their actual behavior depends on how they are mapped onto the underlying infrastructure. This mapping is influenced by CCL decisions, such as collective algorithm selection, channel scheduling, and topology-aware organization, but is ultimately constrained by the physical and logical arrangement of links among accelerators and nodes.

The analysis so far can therefore be summarized as a sequence of connected steps. The model and tokenized input determine quantities such as the parameter count, batch size, sequence length, and activation dimensions. These quantities define the gradient and activation payloads. The parallelization strategy then determines which payloads are exchanged, how often communication occurs, and which accelerators participate. CCO algorithms and CCL implementations further determine how these logical exchanges are divided and scheduled across the network. Because these strategies produce traffic with different sizes, frequencies, and synchronization requirements, they also place different demands on the network. Table~\ref{tab:parallelism_infrastructure_mapping} summarizes this connection and provides a roadmap for the infrastructure discussion that follows.
\begin{table}[t]
\centering
\caption{Relationship between parallelization strategies, communication
patterns, and network infrastructure requirements.}
\label{tab:parallelism_infrastructure_mapping}
\scriptsize
\setlength{\tabcolsep}{4pt}
\begin{tabularx}{\linewidth}{lXXXX}
\toprule
\textbf{Strategy} &
\textbf{Dominant communication} &
\textbf{Traffic timing} &
\textbf{Main network requirement} &
\textbf{Typical placement} \\
\midrule

DP &
Model-sized gradient synchronization, typically through AllReduce. &
Once per optimizer update, after any gradient-accumulation steps. &
High-bandwidth and scalable collective execution for large payloads. &
May span multiple nodes, often using hierarchical intra-node and inter-node
collectives. \\

PP &
Activation and activation-gradient P2P transfers between adjacent pipeline
stages. &
Once per mini/micro-batch at each pipeline boundary in both directions. &
Low latency and enough bandwidth to transfer activations without delaying the next stage. &
Adjacent stages are preferably placed close together or connected through
high-capacity paths. \\

TP/SP &
Frequent activation-sized collectives within Transformer layers. &
Multiple times per layer during both the FW and BW passes. &
Very high-bandwidth and low-latency because communication is frequent and
tightly synchronized. &
Usually kept within a node or another high-bandwidth communication domain. \\

Hybrid &
Combination of the DP, PP, and TP/SP communication patterns. &
The different patterns coexist and may overlap during training. &
Topology-aware placement, balanced link utilization, and contention control. &
TP is generally kept local, PP connects consecutive stages, and DP spans
replicated model or pipeline groups. \\

\bottomrule
\end{tabularx}
\end{table}
The table provides the link between the communication analysis of the previous
sections and the infrastructure discussion that follows. Because these communication patterns place different demands on the network, this section examines the technologies and design choices available to meet them. We discuss interconnects, protocols, topologies, and routing mechanisms, and relate their bandwidth, latency, and scaling properties to the needs of different distributed training configurations. The goal is to provide practical guidance for matching available network
solutions to the communication requirements of DP, PP, TP, and hybrid
parallelism, while accounting for the model and input properties that determine
the size and frequency of the exchanged tensors.

\subsection{Infrastructure hierarchy in AI data centers}
Large scale LLM training is typically deployed within highly integrated AI data centers, where thousands of accelerators can be coordinated to behave as a single training system~\cite{meng2025astral}. From the perspective of this tutorial, the data center provides the physical substrate over which the communication objects identified in the previous sections are transported. Before discussing specific interconnect technologies and protocols, it is useful to distinguish the main layers of this infrastructure and the types of traffic they carry.

\begin{figure}[t]
    \centering
    \includegraphics[width=\linewidth]{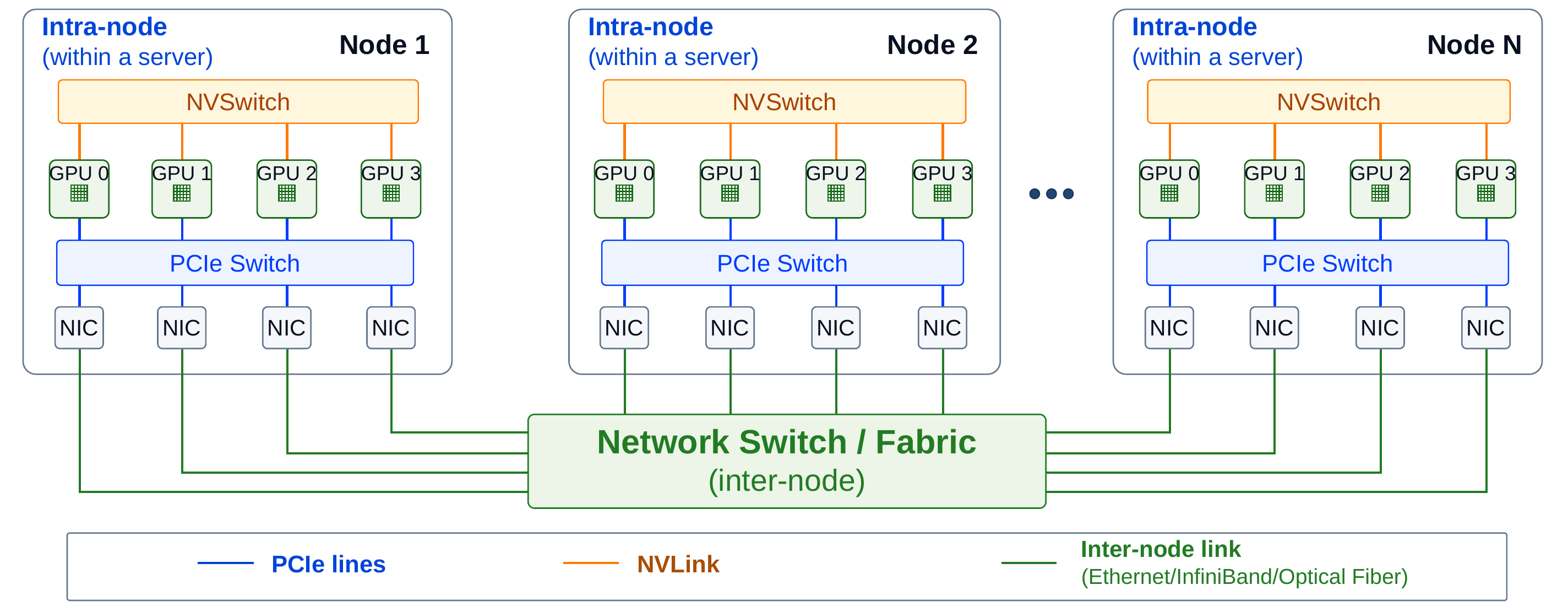}
    \caption{Backend communication hierarchy in distributed LLM training.
    Accelerators within a node communicate through local high-speed
    interconnects, while communication between nodes crosses the data-center
    network fabric.}
    \label{fig:Intra-inter_node}
\end{figure}

A first distinction is between frontend and backend network traffic. The frontend network carries auxiliary traffic associated with the operation of the training system, such as dataset ingestion and checkpoint storage~\cite{zhong2025youmu,wan2025bytecheckpoint}, together with logging, monitoring, and management access. This traffic is often described as \textit{north-south} traffic, since it connects the training cluster with storage systems, external services, or control plane components. In contrast, the backend network carries the communication generated by the training process itself. This traffic is commonly referred to as \textit{east-west} traffic, since it flows among accelerators and nodes participating in the same distributed training job~\cite{qian2024alibaba, gangidi2024rdma}. For large scale LLM training, backend east-west traffic is the dominant concern for performance, as communication time constitute an important percentage of total training time~\cite{narayanan2021efficient}. Fig.~\ref{fig:Intra-inter_node} outlines the backend infrastructure, distinguishing intra-node from inter-node communication and the main interconnects at each level. Accordingly, backend communication can be viewed at two levels:
\begin{itemize}
\item \textbf{Intra-node communication}: this occurs among accelerators located within the same node, where a node can be viewed as a single compute unit composed of one or more CPUs and multiple accelerators interconnected through dedicated onboard interconnects that provide direct device-to-device communication within the server. It supports communication between devices that are physically close and often participate in parallelism groups that require frequent and synchronized data exchange.
\item \textbf{Inter-node communication}: this occurs across different nodes in the training cluster. It enables training jobs to scale beyond a single node capacity and carries communication between nodes over the data center network infrastructure.
\end{itemize}
This hierarchy is important because intra-node and inter-node communication have different performance characteristics and place different requirements on the infrastructure. Intra-node communication typically offers lower latency and higher actual bandwidth, while inter-node communication must scale across many servers and is more exposed to routing, contention, and congestion effects. Because training is synchronized, a small number of slow workers can delay the
entire job even when the communicated tensor payload remains unchanged
~\cite{lin2025understanding}. As a result, the same logical communication operation may behave differently depending on whether it remains inside a node or crosses node boundaries. 
\subsection{Interconnect technologies and protocol mechanisms}
The hierarchy introduced above separates backend training communication into intra-node and inter-node exchanges. These two levels are realized through different interconnect technologies and protocol mechanisms, which affect not only the achievable bandwidth and latency, but also the amount of traffic observed below the payload level. In the previous sections, communication volume was derived from tensor sizes and collective operations. At the infrastructure level, this payload is transported through concrete links and protocols, where packetization, headers, flow control, reliability mechanisms, memory-copy paths, and implementation details determine the actual data movement experienced by the system.

\subsubsection{Intra-node interconnects}
Intra-node communication takes place among accelerators within the same server. A baseline technology for this communication is PCI Express (PCIe), which connects CPUs, GPUs, storage devices, network interfaces, and other accelerator cards inside a node. PCIe communication is organized into lanes, commonly grouped into configurations such as x4, x8, or x16, with modern GPUs typically using x16 connections. For example, PCIe 6.0 can provide a nominal bandwidth of up to approximately 128~GB/s in an x16 configuration. From a traffic characterization perspective, PCIe transports the training payload together with protocol and control information. Data transfers are packetized into Transaction Layer Packets (TLPs), and additional metadata is introduced for reliability, flow control, acknowledgments, ordering, and integrity checks~\cite{sharma2020pci}. Estimating this overhead precisely is difficult because it depends on the PCIe generation, packet size, transfer pattern, and hardware implementation. Direct low-level measurement also typically requires specialized instrumentation. For the purpose of this tutorial, prior studies~\cite{neugebauer2018understanding} and vendor or technical documentation~\cite{altera2018an829, lawley2014wp350} suggest that, for large transfers such as those commonly generated during distributed training, PCIe protocol overhead can be approximated as roughly 5-10\% of the payload traffic. Although PCIe remains an important intra-node communication substrate, it presents limitations for large-scale GenAI training. First, GPU-to-GPU communication over PCIe may share paths with CPUs, memory, storage, and network interfaces, reducing the actual bandwidth available to training traffic. Second, communication across multiple GPUs can be affected by the internal PCIe layout of the server, including switches and CPU domains. These limitations have motivated the adoption of dedicated accelerator interconnects. In NVIDIA-based systems, NVLink provides direct high-bandwidth GPU-to-GPU communication, while NVSwitch extends this design by providing a switching fabric that allows multiple GPUs within a node to communicate with high aggregate bandwidth. These technologies make the node behave more like a tightly coupled accelerator system and are particularly beneficial for communication-intensive parallelization strategies, such as TP. Their detailed low-level overhead is harder to characterize because the implementation is proprietary, but they are designed to provide higher actual bandwidth and lower communication overhead than PCIe-based GPU-to-GPU transfers. The performance characteristics of such intra-node interconnects have been analyzed in prior work, including the study by Li et al.~\cite{li2019evaluating}.
\subsubsection{Inter-node interconnects and Remote Direct Memory Access (RDMA)}
Inter-node communication occurs across different nodes in the training cluster and commonly relies on RDMA-capable high performance fabrics. Traditional TCP/IP communication requires data to traverse the operating system networking stack, often involving multiple memory copies and significant CPU intervention. As link speeds increase to hundreds of gigabits per second, this software mediated path can become a bottleneck for distributed AI training~\cite{kfoury2024comprehensive}. RDMA addresses this limitation by allowing Network Interface Cards (NICs) to transfer data directly between memory regions on different nodes, reducing CPU involvement and communication overhead~\cite{kalia2016design}. In GPU clusters, this idea is extended to accelerator memory. GPUDirect RDMA enables compatible NICs to access GPU memory directly, avoiding intermediate copies through host memory and improving the efficiency of inter-node GPU communication~\cite{potluri2013efficient}.  RDMA is a communication mechanism rather than a single network protocol. One widely deployed implementation is InfiniBand, which provides a native RDMA communication stack and a dedicated high-performance fabric based on InfiniBand adapters, links, and switches~\cite{pfister2001introduction}. RDMA can also be deployed over Ethernet through RDMA over Converged Ethernet (RoCE). RoCE v1 operates directly at the Ethernet link layer and is therefore limited to layer 2 domains. RoCE v2 encapsulates RDMA traffic within UDP/IP packets, enabling layer 3 routing and making RDMA communication possible across routable Ethernet networks~\cite{mittal2018revisiting}. Another RDMA-over-IP approach is iWARP, which runs over TCP/IP, although it is less common in large-scale LLM training deployments~\cite{gross1989communication}.

\subsubsection{Observability and protocol overhead in interconnects}
The choice of interconnect also affects how easily traffic can be measured and how protocol overhead can be estimated. Intra-node technologies such as PCIe, NVLink, and NVSwitch are difficult to inspect directly at the packet or transaction level without specialized hardware support, and proprietary accelerator fabrics expose limited low-level protocol information. In contrast, inter-node traffic carried over Ethernet-based RoCE v2 can often be captured and analyzed using standard packet-inspection tools, since RDMA traffic is encapsulated in UDP/IP packets. This makes header overhead and packet-level behavior easier to study. For the large messages commonly exchanged during distributed LLM training, the protocol overhead of RDMA based inter-node communication is typically small compared with the tensor payload, often limited to a few percent under favorable conditions. However, the actual traffic observed on the network also depends on MTU size, encapsulation, congestion-control behavior, retransmissions, and implementation choices. Therefore, payload-level estimates derived from CCOs should be interpreted as baseline communication requirements, while actual network traffic may include additional protocol and control overheads. Table~\ref{tab:interconnect_summary} summarizes the main interconnect technologies discussed in this section and highlights their role in traffic characterization. The key distinction is between payload traffic, determined by the training strategy and collective operation, and actual network traffic, which also depends on the interconnect, protocol stack, and implementation overheads. The table omits fixed bandwidth and latency values because they depend on hardware generation, device model, link configuration, topology, and deployment. Numerical traffic estimates should instead state their assumptions in a separate parameter table, distinguishing nominal link bandwidth from the actual bandwidth seen by the training workload.
\begin{table}[t]
\centering
\caption{Interconnect Technologies for Distributed LLM Training.}
\label{tab:interconnect_summary}
\begin{tabularx}{\columnwidth}{@{}llX@{}}
\toprule
\textbf{Level} & \textbf{Technology} & \textbf{Role and traffic-characterization implication} \\
\midrule
Intra-node & PCIe & General-purpose server interconnect connecting CPUs, GPUs, NICs, storage, and other devices. GPU-to-GPU transfers may share paths with other components and may depend on the server PCIe layout. Traffic includes protocol overhead due to packetization, ordering, flow control, reliability, and integrity mechanisms. \\
Intra-node & NVLink/NVSwitch & Dedicated accelerator interconnect for high-bandwidth GPU-to-GPU communication within a node. It reduces reliance on PCIe paths and is relevant for communication-intensive parallelism dimensions such as TP. Low-level protocol overhead is difficult to expose because implementations are proprietary.
\\
Inter-node & InfiniBand & Dedicated RDMA-capable fabric widely used in HPC and AI clusters. It provides native RDMA support for low-latency and high-bandwidth node-to-node communication. Traffic characterization must account for message size, congestion, routing, and collective implementation. \\
Inter-node & RoCE v2 & RDMA over routable Ethernet using UDP/IP encapsulation. It enables RDMA communication over Ethernet data center networks and allows packet-level inspection with standard tools. Header overhead is easier to estimate than for proprietary intra-node fabrics, but actual performance depends on MTU, congestion control, loss behavior, and Ethernet fabric configuration. \\
Memory path & GPUDirect RDMA & Mechanism that allows compatible NICs to access GPU memory directly, avoiding intermediate copies through host memory. It reduces CPU involvement and host memory traffic, improving the efficiency of inter-node GPU communication. It affects the memory-copy path rather than defining a separate network fabric. \\
\bottomrule
\end{tabularx}
\end{table}
\subsection{Data center topology and routing effects}

Once training spans multiple nodes, the data center topology determines how the logical communication schedule is mapped onto physical links. Although topology does not change the original tensor payload, it affects the number of traversed hops, available path diversity, bisection bandwidth, contention, and the actual bandwidth observed by the training workload. Consequently, communication groups should be placed according to their traffic characteristics. TP and SP groups generally benefit from high-bandwidth local communication domains; consecutive PP stages should be placed close together; and DP groups require sufficient inter-node bisection bandwidth for large gradient synchronizations.

Modern AI clusters often use multiple network interfaces organized into independent or partially independent \emph{rails}. Rail-aware placement and routing can distribute synchronized collective traffic across these paths, whereas poor assignments may create hotspots despite sufficient aggregate capacity. Equal-cost multipath, adaptive routing, and rail-aware routing similarly influence link utilization. Congestion control, buffering, and loss recovery further affect communication completion times, particularly in Ethernet-based RDMA deployments.

Production systems illustrate these considerations through different designs, including Spectrum-X, Alibaba HPN, large-scale Meta RoCE fabrics, and the optically reconfigurable TPU v4 interconnect ~\cite{nvidia_spectrum_x_2024,qian2024alibaba, gangidi2024rdma,jouppi2023tpu}. Recent work also treats topology as a design variable that can be optimized for communication patterns, scalability, reliability, and cost~\cite{yan2025atop}. Traffic characterization should therefore distinguish logical communication volume from infrastructure-dependent wire-level behavior, which is determined by placement, routing, and contention.
\section{Conclusions and future work}
\label{conclusion}
The growing scope of LLM training has made communication a central component of system performance. Understanding this communication requires linking concepts that are commonly studied in isolation, from datasets and model architectures to distributed execution and network infrastructure. This tutorial introduced the \lq Life of a Token\lq framework to highlight this connection, tracing data from raw text and tokenization through Transformer tensors, parallel execution, collective communication, and finally to the underlying network traffic. The analysis led to several observations that cut across the various levels. Choices regarding the dataset and tokenization determine the number and length of the sequences processed during training, while the model size determines the sizes of the activation, parameter, and gradient tensors. The parallelization strategy determines which tensors become communication objects, which accelerators exchange them, and how frequently these exchanges occur. Collective algorithms then transform the logical payloads of the tensors into communication patterns and into traffic. Finally, the infrastructure determines how these transfers translate into training completion times and scalability. Consequently, no single metric, such as the number of parameters, tensor size, or nominal link bandwidth, is sufficient to characterize the communication behavior of a distributed training configuration. Communication optimization should therefore be considered a cross layer problem. At the software level, 
collective communication libraries can improve the operations 
through algorithm selection, hierarchical execution, message partitioning, scheduling, and overlap with computation. At the infrastructure level, accelerator and network designers continue to increase intra-node and inter-node communication capacity and reduce communication latency. Network topology and accelerator placement also determine which transfers use high-capacity local paths and which traverse the broader fabric. Conversely, model partitioning and hybrid parallelization strategies can be adapted to the available topology; these optimization dimensions are closely interrelated: improving one level without considering the others might simply shift the bottleneck elsewhere. 

Future systems should therefore design the model architecture, parallelization, collective implementation, placement, and network infrastructure in an increasingly integrated manner, optimizing the system as a whole instead of treating each component independently~\cite{wei2024communication,wang2023topoopt}. Rather than prescribing a single optimal configuration, this tutorial provides a framework for identifying where communication originates and where optimization opportunities arise, helping model designers assess infrastructure implications and networking researchers derive traffic requirements from training configurations. Recent work increasingly frames the understanding of LLMs as a multi level scientific problem, spanning learning dynamics, emergent capabilities, internal representations, and the mechanisms that develop during training~\cite{simon2026there, gan2026beyond, zhao2025random, wang2025embryology}. These efforts point toward more predictive and falsifiable explanations of deep learning. This tutorial complements them with a systems perspective, showing how large scale training gives rise to tensor exchanges, parallel execution, collective communication, and network traffic. The \emph{Life of a Token} framework helps make this process more measurable and understandable.

\section*{Acknowledgements}
The authors would like to thank Massimo Gallo for his feedback and title idea. We used generative AI tools to assist with figure design and to correct grammar and spelling.

This research was conducted as part of the Net4AI project, funded by the French National Research Agency (ANR) under contract ANR-24-CE25-5120.

\clearpage
\appendix
\setcounter{secnumdepth}{2}
\refstepcounter{section}
\label{app:extensions}

\section*{Appendix \thesection: Scope and organization}

The main body establishes simplified communication baselines for Data Parallelism (DP), Pipeline Parallelism (PP), and Tensor Parallelism (TP). This appendix extends that analysis in four directions.~\ref{app:architectural_context} relates the original encoder-decoder Transformer to the decoder-only architecture used throughout the tutorial.~\ref{app:dense_refinements} examines how state sharding, pipeline scheduling, and sequence partitioning modify the communication baselines.~\ref{app:ddp_case_study} connects the analytical DP model to measurements from a controlled small-cluster experiment. Finally,~\ref{app:moe} follows token representations through a Mixture-of-Experts model and derives the communication introduced by Expert Parallelism (EP).

Unless stated otherwise, the notation follows the main tutorial. In particular, $B$ is the local mini-batch size, $T$ the sequence length, $d$ the hidden dimension, and $s_{\mathrm{act}}$ the number of bytes used for each activation value. The activation payload is therefore \begin{equation} M = BTd\,s_{\mathrm{act}}. \label{eq:supp_activation_payload} \end{equation} We use FW and BW for the forward and backward passes, CCOs for collective communication operations, and $p_{\mathrm{cc}}$ for the number of accelerators in the relevant communication group.

\subsection{From the Original Transformer to Decoder-Only LLMs}
\label{app:architectural_context}
\begin{figure}[t]
    \centering

    \subfloat[Original encoder--decoder Transformer%
        \label{fig:Transformer_architecture}]{
        \makebox[0.31\linewidth][c]{%
            \includegraphics[
                height=0.28\textheight
            ]{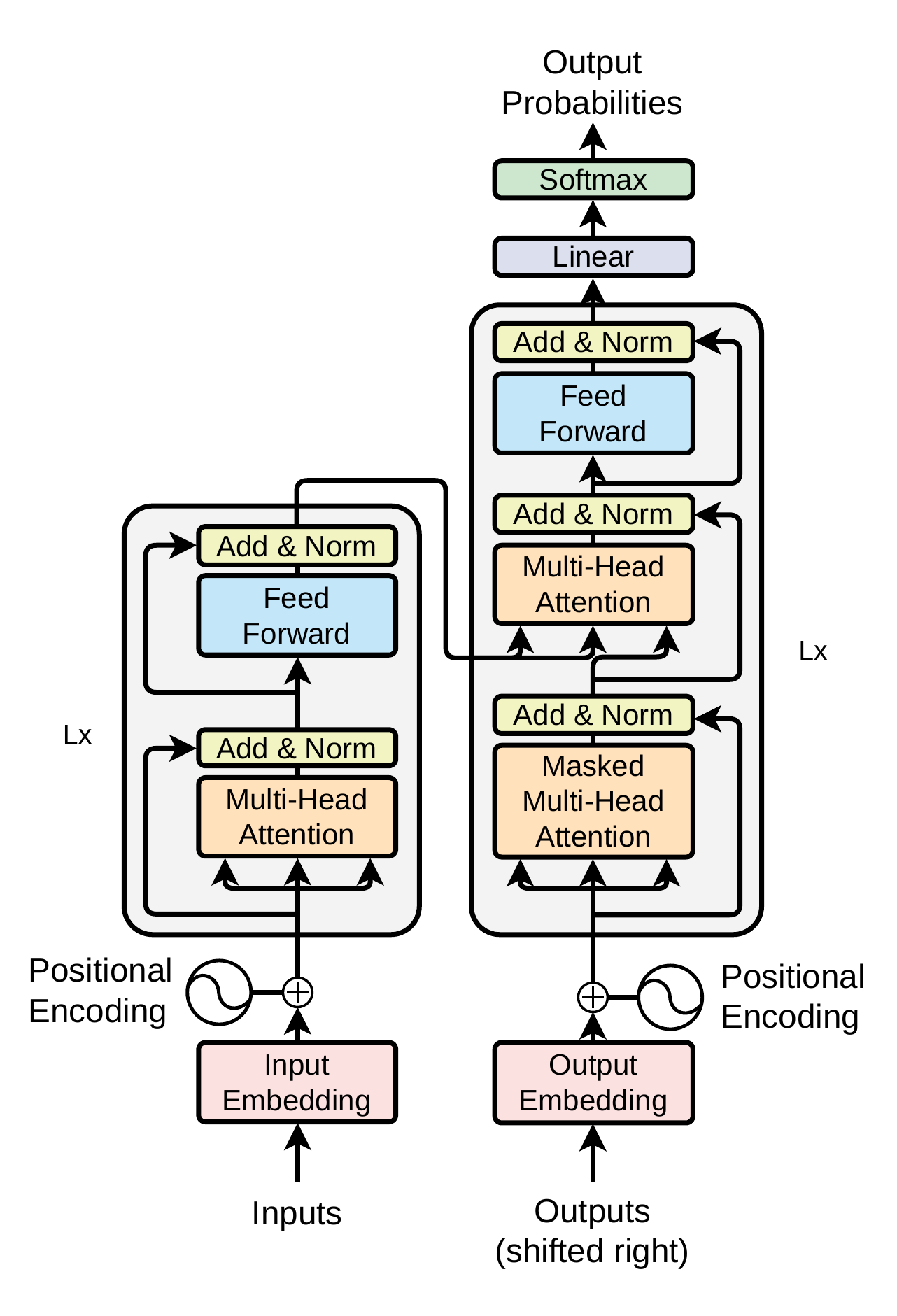}%
        }
    }
    \hfill
    \subfloat[Decoder-only model%
        \label{fig:transformer_decoder_only}]{
        \makebox[0.31\linewidth][c]{%
            \includegraphics[
                height=0.28\textheight
            ]{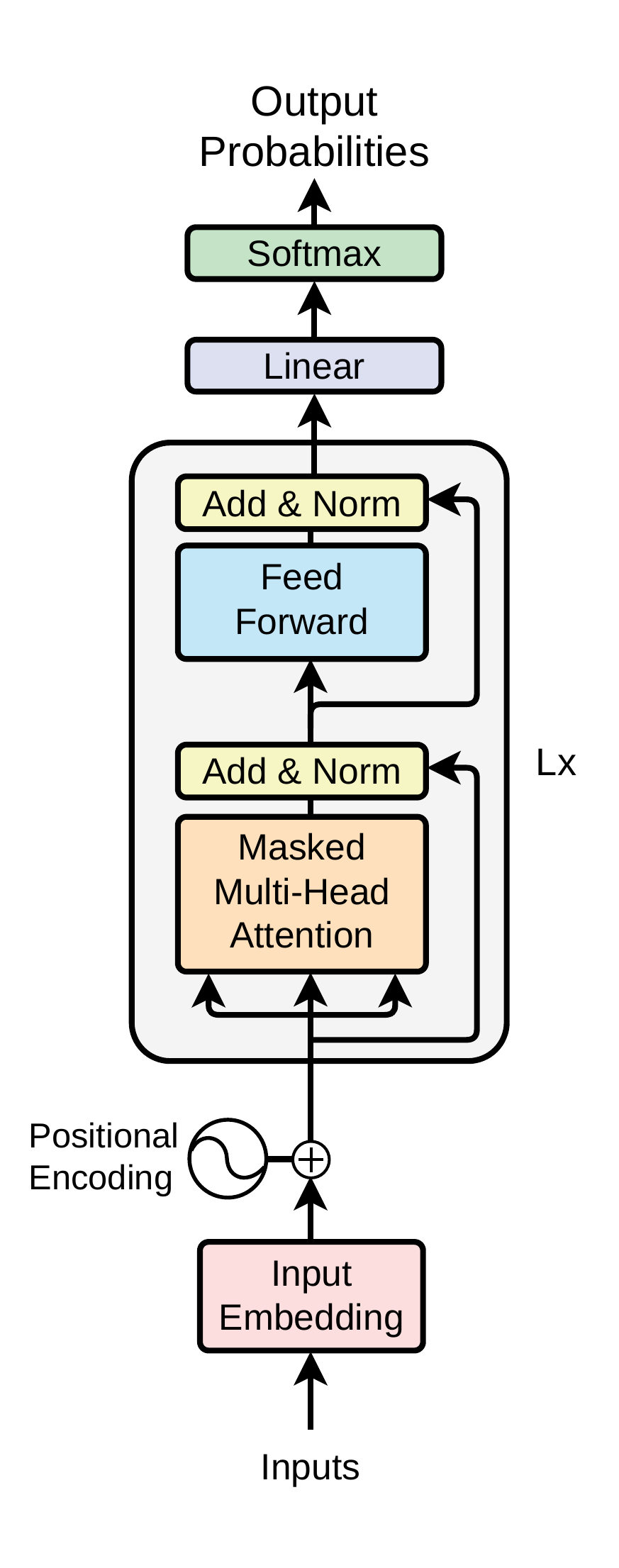}%
        }
    }
    \hfill
    \subfloat[Token processing in an MoE layer%
        \label{fig:moe_layer}]{
        \makebox[0.31\linewidth][c]{%
            \includegraphics[
                height=0.28\textheight
            ]{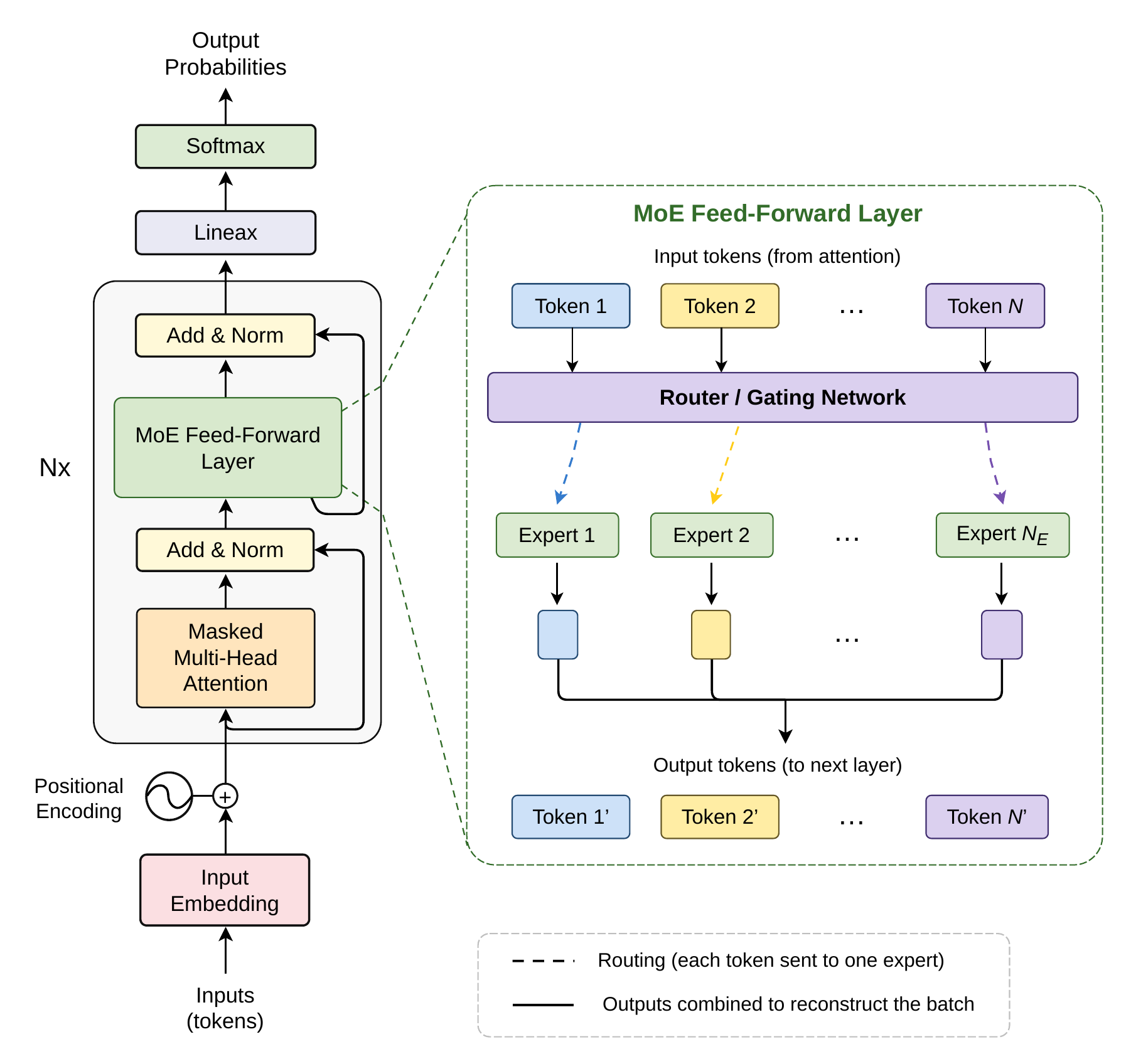}%
        }
    }

    \caption{Comparison of:
    (a) the original encoder--decoder Transformer architecture
    \cite{vaswani2017attention};
    (b) the GPT-2-inspired decoder-only Transformer adopted as the
    reference architecture throughout this tutorial
    \cite{radford2019language}; and
    (c) a representation of a decoder-only Mixture-of-Experts (MoE)
    architecture.}
    \label{fig:transformer_comparison}
\end{figure}

The original Transformer was introduced as an encoder-decoder architecture for sequence-to-sequence tasks such as machine translation~\cite{vaswani2017attention}. As shown in Fig.~\ref{fig:Transformer_architecture}, the encoder processes the source sequence through repeated self-attention and feed-forward blocks. The decoder generates the target sequence using masked self-attention, cross-attention over the encoder output, and a feed-forward network (FFN). The encoder and decoder therefore play distinct roles: the former constructs a representation of the input sequence, while the latter conditions on that representation and on the previously generated target tokens.

Decoder-only models remove the encoder stack and the encoder-decoder cross-attention path. Instead, they apply a single repeated stack of causally masked self-attention and FFN blocks to predict each token from the tokens that precede it. This design was popularized by GPT-style language models~\cite{radford2019language,brown2020language} and constitutes the reference architecture used throughout the main tutorial. Figure~\ref{fig:transformer_comparison} compares the original encoder--decoder architecture in panel (a) with the decoder-only architecture in panel (b). Panel (c) additionally previews the Mixture-of-Experts (MoE) architecture, in which the dense FFN is replaced by multiple experts and token representations are routed to selected experts for processing. This routing process preserves the input and output tensor shape at the layer boundary but introduces specialized communication patterns. The MoE architecture and its communication implications are examined in detail in~\ref{app:moe}

\subsection{Communication Refinements for Dense-Model Parallelism}
\label{app:dense_refinements}

The communication baselines in the main tutorial describe the dominant behavior of each parallelism dimension: model sized gradient synchronization for DP, activation-sized P2P transfers for PP, and frequent activation sized collectives within Transformer layers for TP. Practical systems refine these baselines to reduce memory redundancy, pipeline idle time, or exposed communication. Such refinements may change the collective sequence and message granularity even when the total communicated volume remains of the same order.

\subsubsection{State sharding in data parallelism}
The most widely adopted family of optimizations for DP training focuses on reducing memory redundancy by partitioning optimizer states, gradients, or model parameters across accelerators, instead of fully replicating them on every device. Missing information is reconstructed through CCOs when needed. Two widely adopted examples are ZeRO~\cite{rajbhandari2020zero}, with its variants ZeRO-1, ZeRO-2, and ZeRO-3, and Fully Sharded Data Parallelism (FSDP)~\cite{zhao2023pytorch}. DeepSpeed~\cite{rasley2020deepspeed} is a widely used deep learning optimization library that implements many of these techniques in practice, including the ZeRO family and several of its extensions.

From a network perspective, these approaches progressively redistribute communication across the training process. In standard DP, the ReduceScatter and AllGather phases are executed consecutively as part of a single gradient AllReduce. In ZeRO-1 and ZeRO-2, the overall communication volume remains comparable to standard DP, but the communication schedule is reorganized. In particular, ZeRO-2 partitions gradients, so the ReduceScatter and AllGather phases are separated by the optimizer update rather than being executed as one monolithic gradient AllReduce.

\begin{figure}[t]
    \centering
    \includegraphics[
    width=\linewidth,
    height=0.65\textheight,
    keepaspectratio
    ]{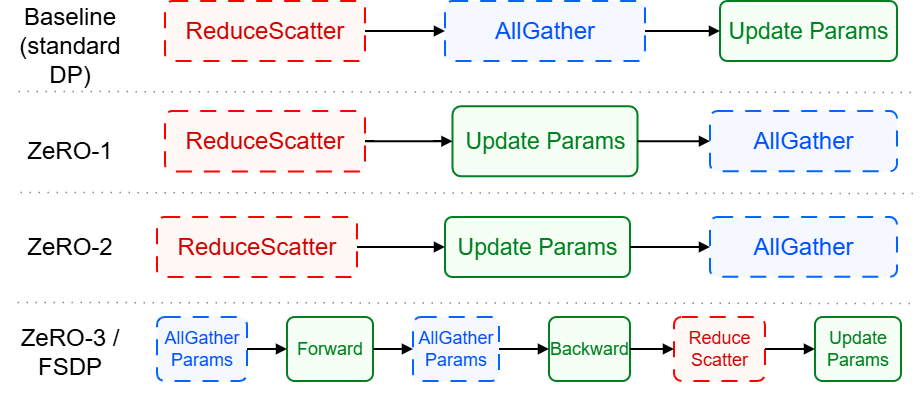}
    \caption{Logical communication schedules in standard DP and ZeRO.}
    \label{fig:DDP_communication_optimizations_schema}
\end{figure}

ZeRO-3 changes the traffic pattern further by also sharding model parameters. Because parameters are no longer fully replicated on each accelerator, the required parameter shards must be gathered before computation. In a simplified view, this introduces an AllGather before the FW pass, another AllGather before the BW pass, and a final ReduceScatter after the BW pass to distribute the reduced gradient partitions. Therefore, unlike ZeRO-1 and ZeRO-2, ZeRO-3 increases the total communication volume because it communicates parameter shards in addition to gradients. The original ZeRO analysis reports this overhead to be approximately \(1.5\times\) the communication volume of standard DP~\cite{rajbhandari2020zero}. If the standard DP traffic is denoted by \(C_{\mathrm{DP}}\), then $C_{\mathrm{ZeRO\text{-}3}} \approx 1.5 C_{\mathrm{DP}}$. FSDP follows a conceptually similar approach by sharding model states across accelerators, while dynamically reconstructing them during execution~\cite{zhao2023pytorch}. Fig.~\ref{fig:DDP_communication_optimizations_schema} summarizes when CCOs are performed in standard DP and in the ZeRO variants.
Several refinements and implementation variants have also been proposed, including ZeRO-Offload, ZeRO-Infinity, and ZeRO++~\cite{ren2021zero,rajbhandari2021zero,wang2024zero++}, which further optimize memory usage and communication efficiency.

\subsubsection{Pipeline schedules and communication overlap}

The main objective of PP optimization is to reduce \emph{pipeline bubbles}. A common approach is to divide each mini-batch into micro-batches. Instead of processing the entire mini-batch at once, different pipeline stages can simultaneously work on different micro-batches, increasing hardware utilization. Early systems such as GPipe~\cite{huang2019gpipe} reduce bubble overhead by splitting mini-batches into many micro-batches and executing them in a pipelined manner. Later approaches, such as PipeDream's one-forward-one-backward (1F1B) scheduling strategy~\cite{harlap2018pipedream}, further improve utilization by overlapping FW and BW passes across different micro-batches.

TeraPipe~\cite{li2021terapipe} further improves pipeline efficiency by introducing token-level PP, where each sequence is split into smaller token slices so that stages can exchange smaller tensors more frequently. More recent works, including Zero Bubble scheduling techniques~\cite{qi2024zero}, aim to further minimize idle periods and improve pipeline efficiency. Another important research direction focuses on communication-aware pipeline execution, where communication between adjacent stages is overlapped with computation in order to reduce the impact of latency and bandwidth limitations. A representative example is DualPipe~\cite{liu2024deepseek}, which explicitly overlaps communication and computation across pipeline stages. Additional optimizations have also been proposed for reducing activation memory usage, preserving weight consistency, and improving the placement and partitioning of pipeline stages across accelerators~\cite{narayanan2021efficient,qi2024pipeline,zheng2022alpa}.

\subsubsection{Sequence and long-context partitioning}
One limitation of the original TP formulation is that only the large matrix multiplications are partitioned across GPUs, while several other operations inside the Transformer layer remain replicated on every accelerator. Although these operations are less computationally intensive, they can still consume a significant amount of memory. To improve memory efficiency, several refinements of TP have been proposed.

One widely adopted extension is sequence parallelism (SP)~\cite{korthikanti2023reducing}, which partitions selected operations along the sequence dimension, reducing memory redundancy while preserving the overall communication structure of TP. From a network perspective, the total communication volume remains broadly comparable to standard TP~\cite{korthikanti2023reducing}, although communication and computation are scheduled differently. As with the PP optimizations discussed previously, SP primarily reorganizes the timing and scheduling of communication rather than fundamentally changing the communication primitives or the amount of exchanged data.

Consequently, the basic TP formulation analyzed above still provides the core intuition required to characterize the resulting network traffic. Once the execution schedule is known, the timing and organization of the communication phases can be derived accordingly. Several additional works have further extended sequence parallelism for long-context LLM training by improving communication efficiency, memory scalability, and execution scheduling across distributed accelerators~\cite{jacobs2023deepspeed,li2023sequence,fang2024usp,liu2024ringattention,li2024distflashattn}.

\subsubsection{Comparative effects of dense-model refinements}
To a first-order approximation, refinements of a given parallelization technique exchange the same underlying tensors as their baseline formulation. The principal differences concern how these tensors are partitioned, when the corresponding communication operations are scheduled, and how they are interleaved or overlapped with computation.

Table~\ref{tab:dense_refinement_summary} relates the resource bottleneck addressed by each refinement to its principal communication effects. This comparison distinguishes the total volume of transferred data from the portion of communication exposed on the execution critical path. Consequently, an optimization may improve E2E throughput without reducing the total byte volume, for example, by dividing transfers into smaller messages, overlapping them with computation, or redistributing them across the training step.

\begin{table}[t]
\centering
\caption{First-order communication effects of dense-model refinements.}
\label{tab:dense_refinement_summary}
\small
\begin{tabularx}{\textwidth}{@{}lXXX@{}}
\toprule
\textbf{Refinement} & \textbf{Primary objective} &
\textbf{Dominant communication} & \textbf{Effect relative to baseline} \\
\midrule
ZeRO-1/2 & Reduce replicated optimizer or gradient state &
ReduceScatter and AllGather around the update &
Comparable first-order volume; schedule is reorganized \\
ZeRO-3/FSDP & Shard parameters, gradients, and optimizer state &
Parameter AllGather and gradient ReduceScatter &
Additional state reconstruction; overlap and bucketing matter \\
PP schedules & Reduce bubbles and exposed transfer time &
Point-to-point activation and activation-gradient transfers &
Similar leading-order bytes; smaller messages and different timing \\
SP and long-context variants & Reduce activation redundancy and support long sequences &
Collectives over sequence or activation partitions &
Implementation-dependent schedule; activation scale remains central \\
\bottomrule
\end{tabularx}
\end{table}

\subsection{Empirical Illustration: DDP Scaling on a Small GPU Cluster}
\label{app:ddp_case_study}

To connect the analytical communication model to E2E behavior, we evaluate a controlled distributed DP workload on a seven-node GPU cluster. Each node contains a dedicated GPU, the nodes are interconnected by 10~Gb/s links, and each GPU hosts one training worker. The workload uses the GPT-2 implementation cited in the main tutorial~\cite{karpathy_build_nanogpt_train_gpt2}.

Communication activity is recorded with \texttt{tcpdump}. The traces are used to identify gradient-synchronization bursts and estimate achieved transfer rates. Table~\ref{tab:ddp_training_results} reports step time, AllReduce time, throughput, and scaling efficiency for one, two, four, and seven workers. 
\begin{table}[t]
\centering
\caption{Measured DDP scaling on the seven-node cluster. Speedup and efficiency
are computed relative to the single-GPU configuration. The AllReduce share is the fraction of
the step time spent in gradient synchronization.}
\label{tab:ddp_training_results}

\small
\renewcommand{\arraystretch}{1.2}
\setlength{\tabcolsep}{6pt}

\begin{tabularx}{\textwidth}{
    @{}
    >{\centering\arraybackslash}X
    >{\centering\arraybackslash}X
    >{\centering\arraybackslash}X
    >{\centering\arraybackslash}X
    >{\centering\arraybackslash}X
    >{\centering\arraybackslash}X
    >{\centering\arraybackslash}X
    @{}
}
\toprule
\textbf{GPUs} &
\textbf{Step time} &
\textbf{AllReduce} &
\textbf{Throughput} &
\textbf{Speedup} &
\textbf{Throughput-scaling efficiency} &
\textbf{AllReduce share} \\
&
\textbf{(ms)} &
\textbf{(ms)} &
\textbf{(tokens/s)} &
\textbf{($\times$)} &
\textbf{(\%)} &
\textbf{(\%)} \\
\midrule
1 & 930  & --  & 1{,}100 & 1.00 & 100 & --   \\
2 & 1{,}130 & 170 & 1{,}810 & 1.65 & 82 & 15.0 \\
4 & 1{,}240 & 250 & 3{,}300 & 3.00 & 75 & 20.2 \\
7 & 1{,}390 & 340 & 5{,}160 & 4.69 & 67 & 24.5 \\
\bottomrule
\end{tabularx}
\end{table}

\subsubsection{Results and interpretation}
Because each worker processes approximately the same number of tokens per step, the global workload increases with the number of workers. The experiment therefore characterizes throughput scaling under an approximately constant per-worker workload rather than strong scaling under a fixed global workload. Aggregate throughput increases from \(1{,}100\) tokens/s with one GPU to \(5{,}160\) tokens/s with seven GPUs, corresponding to a speedup of \(4.69\times\) and a throughput-scaling efficiency of \(67\%\).

As the number of workers increases, the measured AllReduce time rises from \(170\,\mathrm{ms}\) with two GPUs to \(340\,\mathrm{ms}\) with seven GPUs, while its share of the training-step time increases from approximately \(15\%\) to \(24\%\). These measurements indicate that collective communication and synchronization impose increasing overhead under the evaluated configuration, thereby limiting the growth in aggregate throughput. Link bandwidth affects the duration of gradient-synchronization bursts, but collective algorithms, latency, software overhead, stragglers, and the ratio of local computation to communication also influence scaling efficiency. Similar communication-related scaling limitations have been reported in large-scale LLM training studies~\cite{jiang2024megascale,chu2025scaling,go2025characterizing}.

\subsubsection{Limitations and scope}
This experiment provides a controlled empirical illustration of the relationship predicted by the analytical model: as the number of DDP workers increases, gradient synchronization occupies a larger fraction of the training step and reduces scaling efficiency. The seven-node testbed allows this effect to be observed by relating communication time to end-to-end throughput. Although the absolute measurements are specific to the evaluated hardware and 10~Gb/s interconnect and should not be extrapolated directly to modern accelerator clusters, the observed relationship between synchronization cost, communication intensity, and scaling efficiency remains broadly applicable.

\subsection{Mixture of Experts}
\label{app:moe}
The main tutorial focuses on \emph{dense} Transformer models, where every token is processed by the same set of model parameters. One of the most widely adopted modifications to this architecture concerns the FFN block of the Transformer architecture. MoE architectures~\cite{shazeer2017outrageously} replace the single FFN of a Transformer layer with a collection of smaller FFNs, referred to as \emph{experts}. Let $N_E$ denote the number of experts available in a given MoE layer. The key idea is that each token is no longer processed by the same FFN. Instead, after the attention operation, each token is processed only by a specific expert, or by a small subset of experts, selected from the available $N_E$ experts. To determine which expert should process a token, MoE layers employ a small routing network, commonly referred to as a \emph{router} or \emph{gating network}. Given the token representation produced by the attention mechanism, the router computes a score for each expert. The experts with the highest scores are then selected, typically using a top-$K$ strategy, where only the $K$ highest scoring experts are activated for that token. The outputs of the selected experts are subsequently returned to the model's computation and forwarded to the following layers. Fig.~\ref{fig:moe_layer} provides a schematic representation of an MoE layer, showing how tokens are routed to selected experts and subsequently merged to reconstruct the output batch.

The main advantage of MoE architectures is that they increase model capacity without proportionally increasing the computation required to process each token. While the total number of model parameters grows with the number of experts, only a subset of these parameters is activated for a given token. Consequently, MoE architectures can achieve parameter counts significantly larger than those of dense models while maintaining comparable computational requirements.

\begin{figure}[t]
    \centering
    \includegraphics[width=\textwidth]{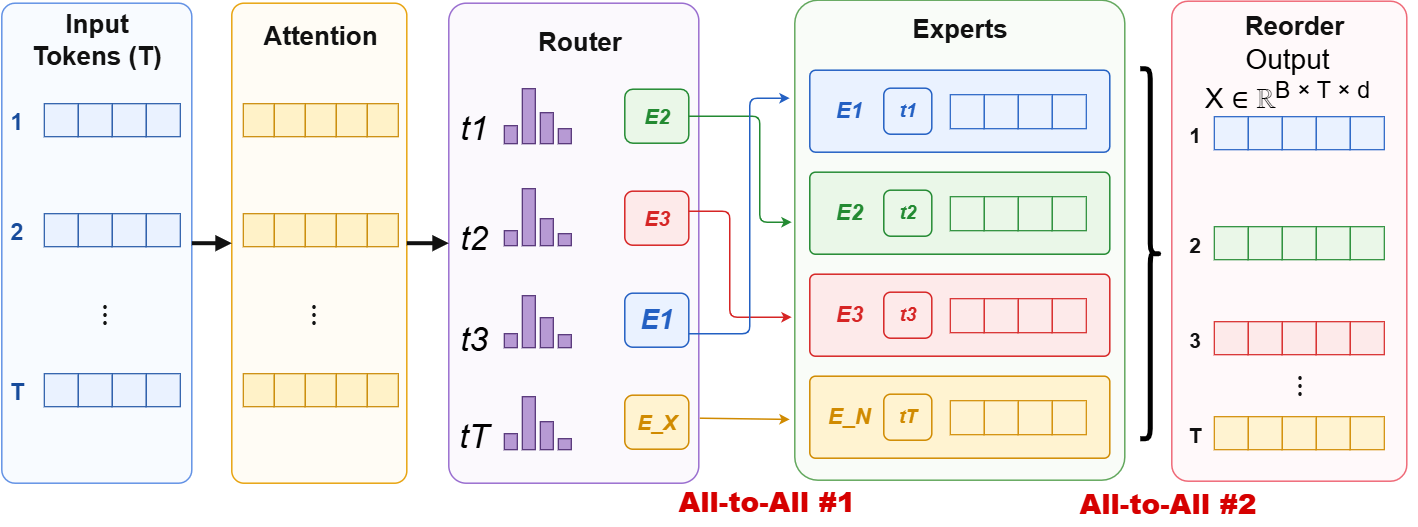}
    \caption{FW pass schema of an MoE Transformer layer.}
    \label{fig:MoE_Tensor_processing}
\end{figure}

To introduce the parallelization strategies commonly adopted for MoE models and understand the communication patterns they generate, it is useful to first examine how a single MoE Transformer layer operates on a single accelerator. Consider an input tensor
$
\mathbf{X} \in \mathbb{R}^{B \times T \times d}.
$
As in a dense Transformer, the tensor is first processed by the attention mechanism, which produces an output tensor of the same dimensions. The routing network then computes a score for each token embedding and assigns it to one or more experts according to the selected routing policy. Consequently, the token embeddings contained in $\mathbf{X}$ are partitioned into groups, with each group processed by a specific expert. After the expert computations are completed, the resulting embeddings are reassembled according to their original token positions, producing an output tensor with the same dimensions as the input tensor. This tensor is then forwarded to the subsequent Transformer layer. During backpropagation, gradients follow the same routing path in the reverse direction. The gradients associated with each token are first propagated through the expert that processed that token and are then reassembled to reconstruct the gradient tensor corresponding to the original input tensor. Fig.~\ref{fig:MoE_Tensor_processing} illustrates how the input tensor is reorganized and processed within a single MoE Transformer layer.

In the next subsection, we examine how MoE models are distributed across multiple accelerators and analyze the resulting communication and network traffic patterns.
\subsubsection{Expert parallelism}
Although MoE architectures increase model capacity without a proportional increase in per-token computation, they still require distributed execution across multiple accelerators. In particular, during training, the complete set of experts must remain resident across the participating accelerators' memories, even though each token activates only a small subset of them. Thus, memory usage scales with the total number of experts, whereas per-token computation depends mainly on the number of activated experts.
The parallelization strategy most commonly associated with is EP~\cite{lepikhin2020gshard, fedus2022switch}. The key idea is to distribute the experts across multiple accelerators while replicating the remaining components of the Transformer layer, such as the attention mechanism and the routing network. During execution, each accelerator processes its local mini-batch in the same way as a dense Transformer up to the routing stage. After the attention operation, the router assigns each token embedding to a specific expert. Since experts are distributed across accelerators, the selected expert may reside on a different device from the one currently processing the token. In this case, the token embedding is transferred to the accelerator hosting the selected expert, then returned after computation and reassembled into the original output tensor before being forwarded to the next Transformer layer.

\begin{figure}[t]
    \centering
    \includegraphics[width=\textwidth]{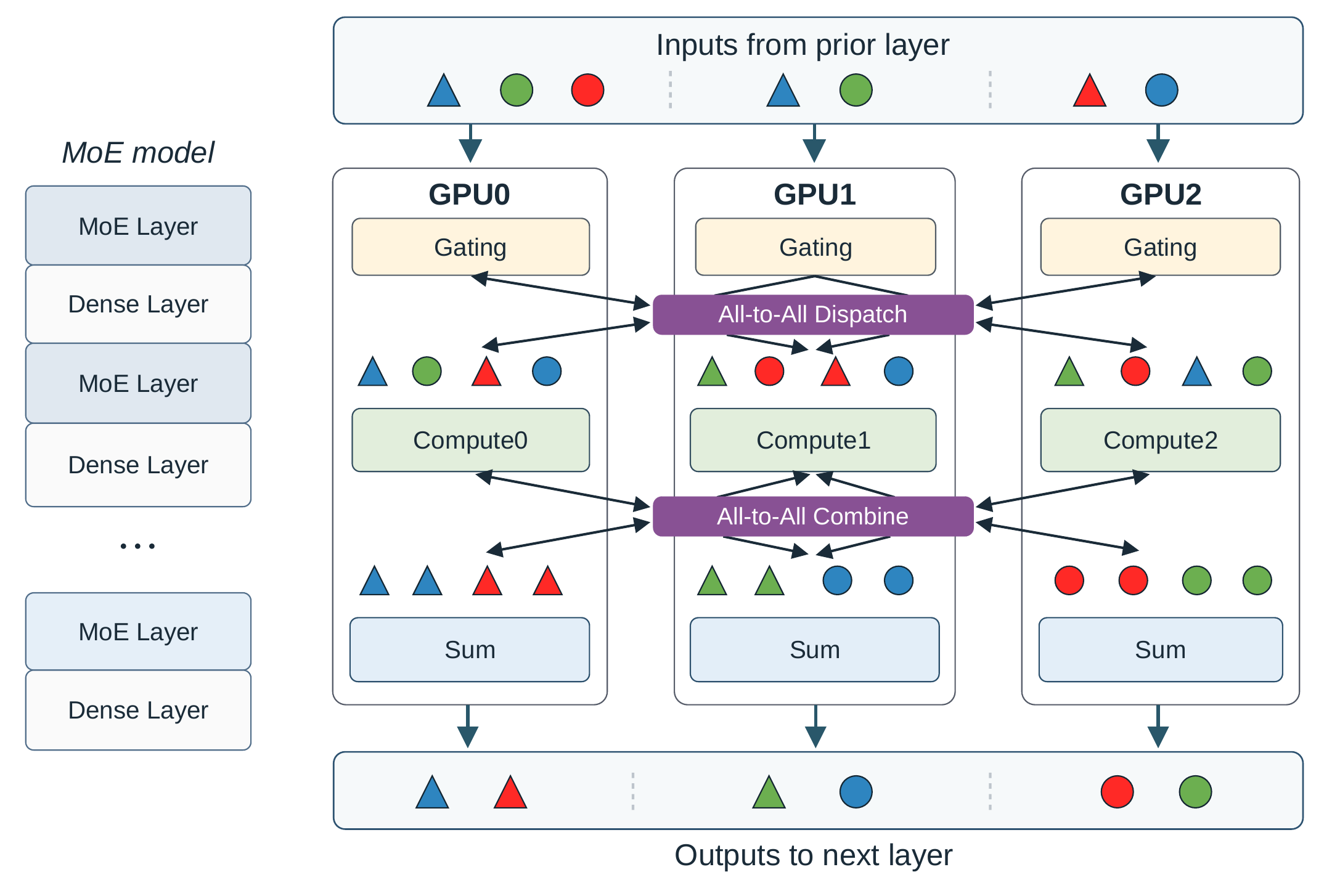}
    \caption{Example of the All-to-All communication pattern adapted from Fig. 2 of~\cite{lei2025flash}.}
    \label{fig:AlltoAll_MoE}
\end{figure}

To maximize hardware utilization, MoE systems typically attempt to balance the number of tokens assigned to each expert. When experts are distributed across multiple accelerators, this routing process generates a communication pattern commonly referred to as \emph{All-to-All}. This communication pattern is typically implemented through the corresponding \emph{All-to-All} CCO. After the routing stage, each accelerator sends the token embeddings assigned to remote experts and receives the embeddings that must be processed by its local experts. Assuming an ideally load-balanced scenario with $p_{\mathrm{cc}}$ accelerators participating in the expert parallel group and a top-\(K\) routing policy, each accelerator exchanges approximately:
$
K \frac{p_{\mathrm{cc}}-1}{p_{\mathrm{cc}}} \mathbf{X},
$
of the local tensor \(\mathbf{X}\). In the common top-1 configuration, \(K=1\), this expression reduces to
$
\frac{p_{\mathrm{cc}}-1}{p_{\mathrm{cc}}} \mathbf{X}.
$ Once the expert computations are completed, a second All-to-All communication phase is performed to return the processed embeddings to their originating accelerators and reconstruct the original tensor organization. Consequently, EP introduces two All-to-All communication phases during the FW pass, with analogous communication occurring during backpropagation. More broadly, MoE training systems can co-schedule intra-node and inter-node communication with expert computation to reduce the exposed All-to-All overhead~\cite{pan2025fsmoe}. Fig.~\ref{fig:AlltoAll_MoE}, adapted from Lei et al.~\cite{lei2025flash} provides a schematic representation of the All-to-All communication pattern introduced by EP.

Let \(M_{\mathrm{MoE}}=BTd\,s_{\mathrm{act}}\) denote the byte payload of the local token-representation tensor. Assuming top-1 routing (\(K=1\)), ideal load balance, and an expert-parallel group of \(p_{\mathrm{cc}}=8\) accelerators, the bytes sent by each accelerator in one All-to-All phase are approximately $C_{\mathrm{A2A}}^{(1)} \approx K\frac{p_{\mathrm{cc}}-1}{p_{\mathrm{cc}}}M_{\mathrm{MoE}} =1\times\frac{8-1}{8}\times50\,\mathrm{MB} =43.75\,\mathrm{MB}.$ Under ideal balance, each accelerator receives the same amount. EP uses two such phases in the forward pass and two analogous phases in the backward pass. Using a sent-bytes convention, the resulting volume is therefore \(4\times43.75=175\,\mathrm{MB}\) per accelerator per MoE layer.

The refinements and sparse-model extension lead to four practical conclusions. First, communicated byte volume and communication exposed on the critical path are different quantities: PP schedules and overlapping techniques may improve throughput without reducing bytes. Second, memory-saving methods can increase or redistribute communication, as illustrated by parameter reconstruction in ZeRO-3 and FSDP. Third, the small-cluster DDP result supports the analytical trend but should be interpreted within its experimental boundary rather than as a modern-cluster benchmark. Finally, EP changes the traffic structure from dense-model gradient or activation exchanges to token-dependent All-to-All communication. Its performance therefore depends not only on tensor size and group size, but also on routing balance, expert placement, and the physical network topology. Together, these observations extend the Life of a Token framework beyond the dense baselines considered in the main tutorial. Whether a token representation is sharded, rescheduled, or routed, the analysis continues to identify the tensor being transferred, the participating devices, the invocation frequency of the communication operation, and the point at which the resulting traffic crosses the infrastructure hierarchy.

\bibliographystyle{elsarticle-num}
\bibliography{biblio}

\end{document}